\documentclass[useAMS,usenatbib]{mn2e}
\usepackage[dvips]{graphicx}
\usepackage{epsfig}
\usepackage{amsmath}
\usepackage{subfigure} 
\usepackage{hyperref}
\title[Identifying merger debris in the {\it Gaia} era]
{On the identification of merger debris in the {\it Gaia} Era} 

\author[F.~A. G\'omez et al.]{Facundo A. G\'omez$^{1}$\thanks{Email:gomez@astro.rug.nl},  Amina Helmi$^{1}$, Anthony G.~A. Brown$^{2}$ \& Yang-Shyang Li$^{1}$\\
$^{1}$Kapteyn Astronomical Institute, University of Groningen, P.O. Box 800, 9700 AV Groningen, the Netherlands\\
$^{2}$Sterrewacht Leiden, P.O. Box 9513, 2300 RA Leiden, the Netherlands
}

\begin{document}

\date{}

\pagerange{\pageref{firstpage}--\pageref{lastpage}} \pubyear{}

\maketitle

\label{firstpage}

\begin{abstract}

We model the formation of  the Galactic stellar halo via the accretion
of satellite galaxies onto a time-dependent semi-cosmological galactic
potential. Our goal is to  characterize the substructure left by these
accretion events  in a close manner  to what may be  possible with the
{\it  Gaia} mission.   We have  created a  synthetic {\it  Gaia} Solar
Neighbourhood catalogue  by convolving the  6D phase-space coordinates
of  stellar particles from  our disrupted  satellites with  the latest
estimates of the {\it Gaia} measurement errors, and included realistic
background  contamination due to  the Galactic  disc(s) and  bulge. We
find  that,  even  after  accounting for  the  expected  observational
errors,  the resulting phase-space  is full  of substructure.   We are
able to successfully isolate  roughly 50\% of the different satellites
contributing to  the `Solar Neighbourhood' by  applying the Mean-Shift
clustering algorithm in  energy-angular momentum space. Furthermore, a
Fourier  analysis of  the space  of orbital  frequencies allows  us to
obtain accurate  estimates of  time since accretion  for approximately
30\% of the recovered satellites.
\end{abstract}


\section{Introduction}

How galaxies form and evolve remains one of the most interesting and
challenging puzzles  in astronomy.  Although a great  deal of progress
has  been  made  in  the  last  decade  many  questions  await  to  be
addressed. Both,  theory and observations  are now converging  about a
key ingredient  of this formation  process; galaxies as our  own Milky
Way seem  to have  experienced the accretion  of smaller  objects that
have  come   together  thanks  to  the  relentless   pull  of  gravity
\citep[e.g.][]{wr}.

From  the observational side,  several new  studies have  revealed the
presence  of a  large amount  of stellar  streams in  galactic stellar
haloes,  echoes of  ancient as  well as  recent and  ongoing accretion
events. These  stellar streams have  been preferentially found  in the
outer   regions   of   galaxies.    The  Sagittarius   Tidal   Streams
\citep{ibata94,ibata01b,majewski} and the Orphan Stream \citep{belu07}
are just two examples of  satellite debris in the Milky Way \citep[see
also][]{new02,ibata03,yanny03,belu06,  grill,  else09}.  The  abundant
substructures    found    not    only    in   the    halo    of    M31
\citep[e.g.][]{ibata01a,McCo09}, but  also in several  haloes of other
galaxies \citep[e.g.][]{Mart08,Mart09}  are an unambiguous  proof that
accretion is inherent to the process of galaxy formation.

On the theoretical side, models  of the formation of stellar haloes in
the  $\Lambda$CDM cosmogony  have  been able  to  explain their  gross
structural properties \citep[e.g.][]{bj05,lh,cooper}. In these models,
the inner  regions of the  haloes (including the  Solar Neighbourhood)
typically formed  first and hence  contain information about  the most
ancient   accretion   events   that   the   galaxy   has   experienced
\citep[e.g.][]   {ws2000,hws03,bj05,tumli}.   Previous   studies  have
predicted  that several  hundreds kinematically  cold  stellar streams
should  be present  in  the Solar  Vicinity  \citep{hw}. However,  the
identification  of  these  streams  is  quite  challenging  especially
because of the  small size of the samples of  halo stars with accurate
3-D velocities  currently available.  Full  phase-space information is
necessary because  of the very  short mixing time-scales in  the inner
regions of the halo. Some progress has been made towards building such
a  catalogue but  only a  few detections  have been  reported  to date
\citep{hwzz99,klement09,smith09}.  These have  made use  of catalogues
such   as  SEGUE   \citep{yanny09}  and   RAVE   \citep{zwitter08}  in
combination  with  Tycho   \citep{thyco}  and  Hipparcos  \citep{hip}.
Clearly, this field will only advance significantly with the launch of
the astrometric satellite {\it  Gaia} \citep{gaia}.  This mission from
the European  Space Agency will  provide accurate measurements  of the
6-D  phase-space coordinates  of  an extraordinarily  large number  of
stars\footnote{Details about the latest {\it Gaia} performance numbers
may  be   found  at:  http://www.rssd.esa.int/gaia}.    Together  with
positions, proper  motions and parallaxes  of all stars  brighter than
$V=20$,  {\it Gaia}  will also  measure  radial velocities  down to  a
magnitude of $V=17$.

To  unravel  the accretion  history  of  the  Milky Way  requires  the
development  of  theoretical  tools  that will  ultimately  allow  the
identification and characterization of the substructure present in the
{\it Gaia} data  set. Several authors have studied  the suitability of
various  spaces to  isolate debris  from accretion  events \citep[see,
e.g.][]{hz00,knebe05,ari_fu,font,h06,sj09}.  Recently \citet[][]{mcm},
followed  by  \citet[][hereafter   GH10]{gh10},  showed  that  orbital
frequencies form  a very  suitable space in  which to  identify debris
from past merger events. In frequency space individual streams from an
accreted  satellite can  be  easily identified,  and their  separation
provides a  direct measurement of its time  of accretion. Furthermore,
GH10 showed  for a few  idealized gravitational potentials  that these
features are preserved  also in systems that have  evolved strongly in
time.  While  very promising, these  studies have focused on  a single
accretion  event  onto a  host  galaxy.   In  reality, we  expect  the
Galactic stellar  halo to have formed  as a result  of multiple merger
events  \citep[where most  of its  mass  should have  originated in  a
handful  of significant  contributors; see  e.g.][]{lh,cooper}.   As a
consequence, it  is likely that debris from  different satellites will
overlap in frequency space, complicating their detection.

In  this  paper  we  follow   multiple  accretion  events  in  a  live
(cosmologically motivated) galactic  gravitational potential. We focus
on how the latest estimates of the measurement errors expected for the
{\it Gaia}  mission will affect  our ability to recover  these events.
In  particular  we  study  the  distribution  of  stellar  streams  in
frequency  space  and  the  determination of  satellite's  time  since
accretion from  stellar particles located  in a region like  the Solar
Neighbourhood.  Our  paper is organized  as follows.  In Section  2 we
present the details  of the $N$-body simulations carried  out, as well
as the  steps taken to generate  a Mock {\it  Gaia} stellar catalogue.
In Section  3 we  characterize the distribution  of debris in  a Solar
Neighbourhood-like sphere in the  absence of measurement errors, while
in  Section  4  we focus  on  the  analysis  of  the Mock  {\it  Gaia}
catalogue. We discuss and summarize our results in Section 5.

\section{Methods}

We model  the formation  of the  stellar halo of  the Milky  Way using
$N$-body  simulations  of the  disruption  and  accretion of  luminous
satellites  onto  a  time-dependent  galactic potential.  We  describe
firstly  the characteristics  of  the galactic  potential  and of  the
population of satellites, as well as the $N$-body simulations. Finally
we  outline the  steps followed  to generate  a mock  {\it  Gaia} star
catalogue.

In this paper we adopt a  flat cosmology defined by $\Omega_{m} = 0.3$
and $\Omega_{\Lambda}=0.7$ with a Hubble constant of $H(z = 0) = H_{0}
= 70$ km~s$^{-1}$Mpc$^{-1}$.

\subsection{The Galactic potential}

\label{pots}

\begin{figure}
\centering
\includegraphics[width=82mm,clip]{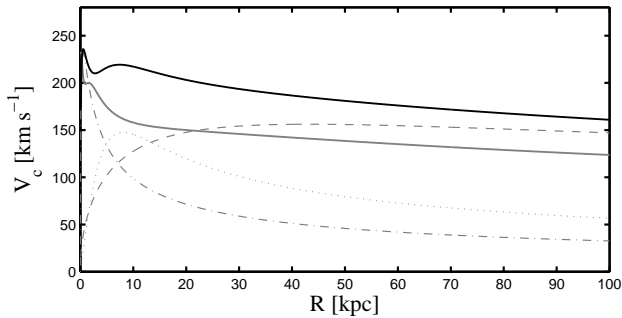}
\caption{Circular velocity  curve as a  function of distance  from the
  galactic centre. The thick  solid lines represent the total circular
  velocity used in  the suite of simulations at ${\it  z} = 0$ (black)
  and  ${\it  z}  \sim  1.85$ (grey),  respectively.   The  individual
  contributions from the  dark matter halo, the disc  and the bulge at
  present time are shown by  the thin dashed, dotted and dashed dotted
  lines, respectively.}
\label{fig:rot_cruve}
\end{figure}

To model  the Milky Way  potential we chose a  three-component system,
including a Miyamoto-Nagai disc \citep{mi-na}
\begin{equation}
\Phi_{\rm disc}=-\frac{GM_{\rm d}}{\sqrt{R^{2}+(r_{a}+\sqrt{Z^{2}+r_{b}^{2}})^{2}}},
\end{equation}
a Hernquist bulge \citep{hernq},
 \begin{equation}
\Phi_{\rm bulge}=-\frac{GM_{\rm b}}{r+r_{c}},
\end{equation} 
and NFW dark matter halo \citep{nfw} 
\begin{equation}
\Phi_{\rm halo}=-\frac{GM_{\rm vir}}{r\left[\log(1+c)-c/(1+c)\right]}\log\left(1+\frac{r}{r_{s}}\right).
\end{equation}
\begin{table}
\begin{minipage}{90mm} 
\caption{Parameters of  the present day Milky  Way-like potential used
  in our $N$-body simulations. Masses are in M$_{\odot}$ and distances
  in kpc.}
\label{table:model}
\begin{tabular}{@{}lll} \hline Disc & Bulge & Halo\\
\hline $M_{\rm d} = 7.5 \times 10^{10}$ & $M_{\rm b} = 2.5 \times 10^{10}$ & $M_{\rm vir} = 9 \times 10^{11}$ \\ $r_{\rm a} = 5.4$; $r_{\rm b} = 0.3$& $r_{\rm c} = 0.5$ & $r_{\rm vir} = 250$ \\  & & $c = 13.1$ \\
\hline
\end{tabular}
\end{minipage}
\end{table} 
Table~\ref{table:model}  summarizes   the  numerical  values   of  the
parameters at  redshift $z  = 0$ \citep{smith07,sofue}.   The circular
velocity curve  in this  model takes a  value of $220$  km~s$^{-1}$ at
$8$~kpc    from   the    galactic    centre,   and    is   shown    in
Figure~\ref{fig:rot_cruve}.

To  model  the evolution  of  the Milky  Way  potential  we allow  the
characteristic parameters to  vary in time. For the  dark matter halo,
the evolution of  its virial mass and its  concentration as a function
of redshift are given by \cite{wech} and \cite{zhao}
\begin{equation}
M_{\rm vir}(z)=M_{\rm vir}(z = 0) \exp(-2a_{c}z),
\end{equation}
where the constant $a_{c} = 0.34$ is defined as the formation epoch of
the halo, and
\begin{equation}
c(z)=\frac{c(z = 0)}{1 + z}.
\end{equation}

For the disc and bulge components we follow the prescriptions given by
\citet{bj05}, i.e.,
\begin{equation}
M_{{\rm d, b}}(z)=M_{\rm d, b}(z = 0) \frac{M_{\rm vir}(z)}{M_{\rm vir}(z = 0)}
\end{equation}
for the masses and
\begin{equation}
r_{\rm a, b, c}(z) = r_{\rm a, b, c}(z = 0) \frac{r_{\rm vir}(z)}{r_{\rm vir}(z = 0)}
\end{equation}
for the scalelengths.  Here $r_{\rm vir}$ is the  virial radius of the
dark matter halo. Its evolution can be expressed as
\begin{equation}
r_{\rm vir}(z) = \left(\dfrac{3 M_{\rm vir}(z)}{4 \pi \Delta_{\rm vir}(z) \rho_{c}(z)}\right)^{1/3}
\end{equation}
where $\Delta_{\rm vir}(z)$ is the
virial overdensity \citep{bn},
\begin{equation}
\Delta_{\rm vir}(z) = 18\pi^{2}+82[\Omega(z)-1]-39[\Omega(z)-1]^{2}
\end{equation}
with $\Omega(z)$ the mass density of the universe,
\begin{equation}
\Omega(z) = \dfrac{\Omega_{m}(1+z)^{3}}{\Omega_{m}(1+z)^{3}+\Omega_{\Lambda}},
\end{equation}
and $\rho_{c}(z)$ is the critical density of the universe at a given
redshift,
\begin{equation}
\rho_{c}(z) = \dfrac{3H^{2}(z)}{8 \pi G},
\end{equation}
with 
\begin{equation}
H(z) = H_{0} \sqrt{\Omega_{\Lambda} + \Omega_{m} (1+z)^{3}} 
\end{equation}

\subsection{Satellite galaxies}

\subsubsection{Internal properties}

\label{sec:int_prop}
\begin{figure}
\centering
\includegraphics[width=82mm,clip]{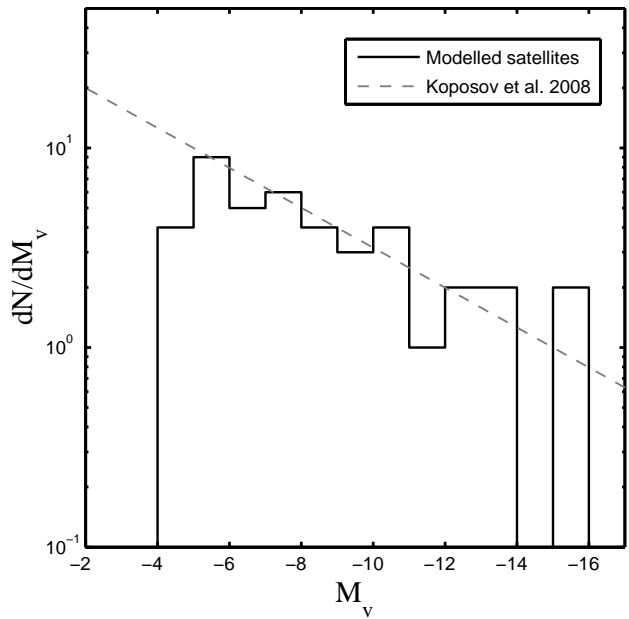}
\caption{Luminosity function for the  model satellite galaxies used in
  this suite of simulations  (histogram). The (grey) dashed line shows
  the `all sky SDSS' luminosity function by \citet{kopo}.}
\label{fig:lum_fun}
\end{figure}

We assume the properties of  the progenitors of our model stellar halo
(i.e. the satellites at high redshift) are similar to those at present
day.   The Galactic  stellar  halo  has a  total  luminosity of  $\sim
10^{9}$ {\it  L}$_\odot$.  To obtain a population  of satellites that,
after accretion,  produces a stellar halo with  this total luminosity,
we  use  the  luminosity   distribution  function  of  the  Milky  Way
satellites given by \citet{kopo}
\begin{equation}
\frac{dN}{dM_{V}}=10 \times 10^{0.1(M_{V}+5)}.
\end{equation}
Figure~\ref{fig:lum_fun}  shows the luminosity  function of  our model
satellite population obtained by randomly drawing 42 objects.

For the initial stellar structure  of the satellites, we assume a King
profile \citep{king} with a  concentration parameter $c \approx 0.72$.
We derive the  structural parameters for each of  our satellites using
the scaling relations
\begin{equation*}
\displaystyle
\begin{array}{ll}
\label{in_freqs}
\displaystyle
\log \frac{L}{L_{\odot}} - 3.53 \log \frac{\sigma_{v}}{\text{km s}^{-1}} \approx 2.35,\\
\\
\displaystyle
\log \frac{\sigma_{v}}{\text{km s}^{-1}} - 1.15 \log \frac{R_{c}}{\text{kpc}} \approx 1.64,
\\
\displaystyle
\end{array}
\end{equation*}
as  given by  \citet{guzman} for  the Coma  cluster  dwarf ellipticals
galaxies.  Here  $R_{c}$ is the core  radius of the  King profile, and
$\sigma_{v}$  is the central  velocity dispersion.   We note  that the
total mass of the King model as defined by these parameters may differ
from  that implied by  the satellite's  luminosity $L$.   Therefore we
also  implicitly  allow the  presence  of  dark  matter in  our  model
satellites.

\subsubsection{Orbital properties}
\label{sec:orb_prop}

The density profile of the stellar halo is often parametrized in a principal
axis coordinate system as

\begin{equation}
\displaystyle
\rho(x,y,z) = \rho_{0} \frac{\left(x^{2}+\dfrac{y^{2}}{p^{2}}+\dfrac{z^{2}}{q^{2}}+a^{2}\right)^{n}}{r_{0}^{n}}
\end{equation}
where  $n$ is  the  power-law exponent,  $p$  and $q$  the minor-  and
intermediate-to-mayor axis ratios, $a$ the scale radius and $\rho_{0}$
the  stellar halo  density  at a  radius  $r_{0}$.  \citep[See][for  a
complete review of  recent measurements of these parameters.]{hreview}
Here for simplicity, we shall assume  $a=0$ (as the halo appears to be
rather concentrated), $p=q=1$ and $n=-3$.  At the solar radius, $r_{0}
= R_{\odot}=8$~kpc,  $\rho_{0}$ corresponds to the  local stellar halo
density.   In this  paper  we  adopt $\rho_{0}  =  1.5 \times  10^{4}$
M$_{\odot}$~kpc$^{-3}$, as given by \citet{fuchs}.

From this density profile we  randomly draw positions at redshift $z =
0$ for  the guiding orbit of  each one of our  satellites. To generate
their  velocities we  follow the  method described  in section  2.1 of
\citet{hz00}. The main assumption made is that the stellar halo can be
modelled by a  radially anisotropic phase-space distribution function,
which is only a function of energy, $E$, and angular momentum, $L$.

This combined set of orbital  initial conditions guarantees that at $z
= 0$, the observed stellar halo density profile and velocity ellipsoid
at the  solar radius are roughly  matched by our  model. These initial
conditions are then integrated backwards  in time for $\sim 10$~Gyr ($
z  \approx  1.85$)  in   the  time-dependent  potential  described  in
Section~\ref{pots}.   Therefore,  we  now  have obtained  the  set  of
initial positions  and velocities required for  forward integration in
time of our $N$-body simulations.

We do not include any numerical treatment of dynamical friction in our
simulations. Therefore, and  to mimic the effects of  this process, we
have assigned the  five most bound orbits at present  time to the five
most massive satellites. For the rest of the satellites the orbits are
assigned randomly.

\subsubsection{Numerical treatments}

The  numerical  simulations  are,  in  most  cases,  self-consistently
evolved   using  the  massively   parallel  TREESPH   code  GADGET-2.0
\citep{springel2005}.  The  number of particles used  to simulate each
satellite  depends  on  its  total  luminosity,  as  explained  bellow
(Section~\ref{sec:mock_cat}).   This number ranges  from a  minimum of
$2.56 \times  10^{5}$ up to  $10^{7}$ particles.  To avoid  very large
computational times we neglect  self-gravity for the five most massive
satellites.  We do not expect this to significantly affect our results
since  these satellites  inhabit the  inner regions  of the  host (see
Section~\ref{sec:orb_prop}) and therefore  they are rapidly disrupted.
For  the  gravitational  softening  parameter we  choose  $\epsilon  =
0.025R_{c}$ (where $R_{c}$  is the core radius of  the system).  After
allowing each satellite  to relax in isolation for  $3$~Gyr, we launch
it from the apocentre of its orbit and let it evolve for $\sim 10$~Gyr
under the influence of the time-dependent galactic potential.

\subsection{Generating a mock {\it Gaia} catalogue}

\label{sec:mock_cat}

\subsubsection{$N$-body particles to stars}

To generate  a mock  {\it Gaia} catalogue  we need to  `transform' our
$N$-body  particles  into   `stars'.   Therefore  we  assign  absolute
magnitudes,  colours  and  spectral  types  to each  particle  in  our
simulations.   We   use  the  IAC-STAR   code  \citep{apagalla}  which
generates  synthetic colour-magnitude  diagrams (CMDs)  for  a desired
stellar population model. As output  we obtain the $M_{V}$ and $(V-I)$
colour for each stellar particle.

\begin{figure}
\centering
\includegraphics[width=82mm,clip]{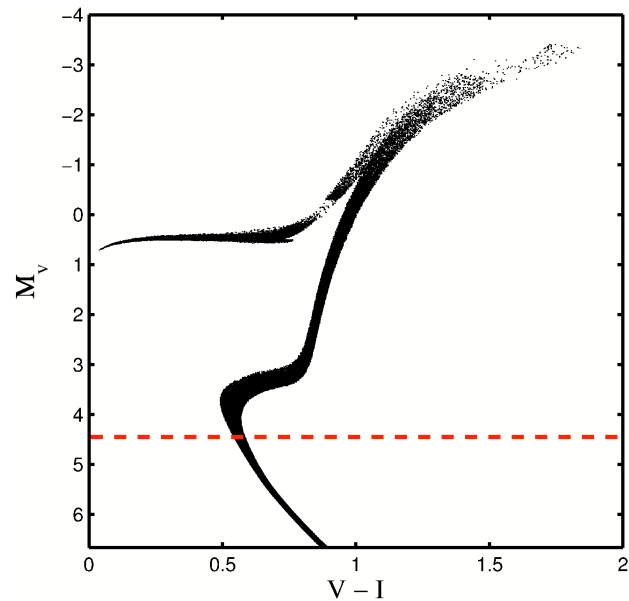}
\caption{Example of a synthetic CMD used to populate our model
  satellite galaxies with `stars'. The dashed line indicates the
  limiting magnitude considered in our simulations ($M_{V} \approx
  4.5$).}
\label{fig:cmd}
\end{figure}

We assign  to each  satellite a  stellar population
model \citep{pietri} with  a range of ages from  $11$ to $13$~Gyr, $-2
\leq [{\rm Fe}/{\rm H}] \leq -1.5$ dex, $[\alpha/{\rm Fe}] = 0.3$, and
a Kroupa  initial mass function \citep{kroupa}.   All these parameters
are consistent with  the observed values in the  Galactic stellar halo
\citep[][and   references  therein]{hreview}.    An  example   of  the
resulting CMD is shown in Figure~\ref{fig:cmd}

Due to  the limited resolution of  our $N$-body simulations  it is not
possible to fully populate the CMDs of our satellites.  Therefore each
satellite is  populated only with  stars brighter than  $M_{V} \approx
4.5$, which corresponds  to an apparent magnitude $V  = 16$ at $2$~kpc
from the  Sun.  This choice  represents a good compromise  between the
numerical  resolution of  our  simulations and  the limiting  apparent
magnitude  $V_{\rm lim}$  for which  {\it Gaia}  will measure  full 6D
phase-space  coordinates  (astrometric  and  photometric  measurements
extend to  $V = 20$,  but spectroscopic measurements reach  $V \approx
17$).

As explained  in Section~\ref{sec:int_prop} different  satellites have
different total  luminosities and, thus,  the number of stars  down to
magnitude $M_{V} \approx 4.5$ varies from object to object. Therefore,
in  each  satellite  a  different  number  of  $N$-body  particles  is
converted into stars.  To estimate this  number we use the CMD code by
\citet{marigo},   which  provides   a   stellar  luminosity   function
normalized to the initial stellar mass of a given population.  The age
and  metallicity of  the  stellar  population model  is  fixed to  the
average  values described above  and the  initial mass  is set  by the
total luminosity  of each satellite. Finally,  accumulating the number
of stars down to $M_{V} \approx  4.5$ we obtain the number of $N$-body
particles that  need to be  transformed into stars. This  number range
from  660 for the  faintest satellite  ($M_{V} =  -5$) to  $1.4 \times
10^{7}$ for the brightest one ($M_{V} = -15.9$).

\subsubsection{Disc and Bulge contamination}

The {\it  Gaia} catalogue will contain  not only stars  from the halo,
but also stars  associated to the different components  of the Galaxy.
These stars act  as a {\it smooth} Galactic  background which could in
principle complicate  the task of identification  of debris associated
to accretion  events.  To take  into account this  stellar background,
\citet[][hereafter BVA05]{brown05} created a  Monte Carlo model of the
Milky Way. This model consists  of three different components: a disc,
a bulge, and a smooth stellar halo; where the latter was meant to take
into account  the set of halo  stars formed in-situ.  In  this work we
only  consider the  background  contamination by  the  disc and  bulge
components of the Monte Carlo  Galactic model of BVA05, since we build
our stellar halo completely from disrupted satellites.  The disc model
consists  of particles  distributed  in space  according  to a  double
exponential  profile.    For  the   bulge  model  a   Plummer  density
distribution \citep[][]{plum} is adopted. To each particle an absolute
magnitude and spectral class is assigned according to the Hess diagram
listed in  table~4-7 of \citet{mihabin}.  Finally,  the kinematics are
modelled assuming different velocity  ellipsoids for stars in the disc
of  each spectral  type whereas  an isotropic  velocity  dispersion is
assumed for bulge stars. For more details, see section~2.1 of BVA05.

\subsubsection{Adding {\it Gaia} errors}

\begin{figure}
\includegraphics[width=85mm,clip]{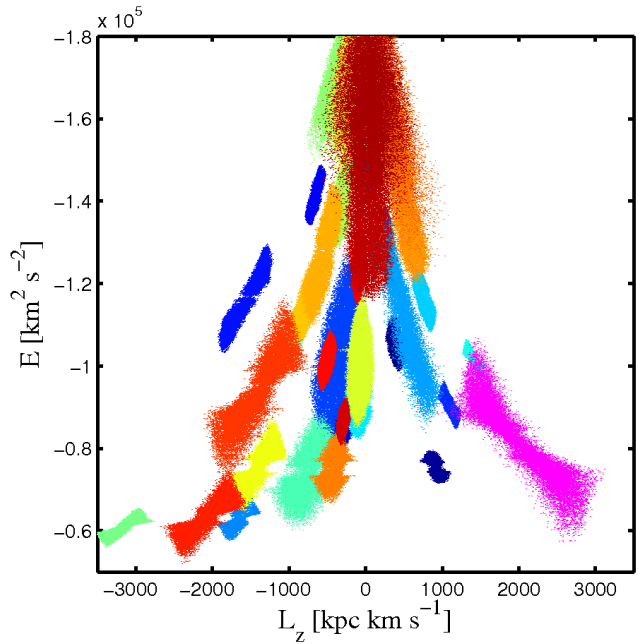}
\caption{Distribution of particles in  $E-L_{z}$ space after 10~Gyr of
  evolution.  Different colours represent different satellites.}
\label{fig:ELz}
\end{figure}

To  convolve the  positions and  velocities of  our mock  catalogue of
stars with the  expected {\it Gaia} errors we  have followed the steps
described in section~4.1  of BVA05.  Here we give  a short overview of
the procedure but we refer the reader to BVA05 for more details.  As a
first step,  we transform  Galactocentric positions and  velocities to
heliocentric  coordinates,  $ \vec{r}_{\rm  helio}  = \vec{r}_{gal}  -
\vec{r}_{\odot}$   and   $\vec{v}_{\rm   helio}  =   \vec{v}_{gal}   -
\vec{v}_{\odot}$. Subsequently, these  quantities are transformed into
radial velocity and five  astrometric observables that {\it Gaia} will
measure,  i.e.,   the  Galactic  coordinates   $(l,b)$,  the  parallax
$\varpi$,  and  the  proper  motions  $\mu_{l*}=\mu_{l}  \cos{b}$  and
$\mu_{b}$.   The next step  consists in  convolving with  the expected
{\it Gaia}  errors according to  the accuracy assessment  described in
the {\it  Gaia} web pages at  ESA (http://www.rssd.esa.int/gaia). Note
that the  errors vary  most strongly with  apparent magnitude,  with a
weaker  dependence on  colour and  position on  the sky.   Finally, we
transformed  the particles' observed  phase-space coordinates  back to
the Galactocentric reference frame for our analysis.

\section{Characterization of satellite debris}

In this  section we  analyse how  the debris of  our 42  satellites is
distributed in various projections  of phase-space, before taking into
account how the {\it Gaia} observations will affect this distribution.
In comparison to previous  works \citep[e.g.][]{hz00}, recall that our
satellites have evolved in a live potential for 10~Gyr.

\subsection{Traditional spaces}

Figure~\ref{fig:ELz}  shows  the  distribution  of  $\sim  2.5  \times
10^{4}$ randomly chosen particles from  each satellite in the space of
energy, $E$, and the $z$-component of the angular momentum, $L_z$. The
different colours  represent different  satellites.  Note that  in this
projection  of phase-space  a large  amount of  substructure  is still
present,  despite  the  strong  evolution of  the  host  gravitational
potential (e.g. the total mass increased by a factor $\sim 3.5$).

\begin{figure}
\includegraphics[width=85mm,clip]{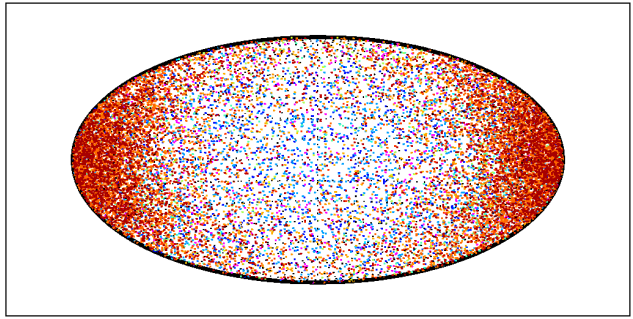}
\caption{Distribution on the sky ({\it l,b}) of the stellar particles
  located inside a sphere of 2.5~kpc radius centred at 8~kpc from the
  galactic centre. Different colours represent different
  satellites.}
\label{fig:mapas}
\end{figure} 

\begin{figure}
\begin{centering}
\includegraphics[width=80mm,clip]{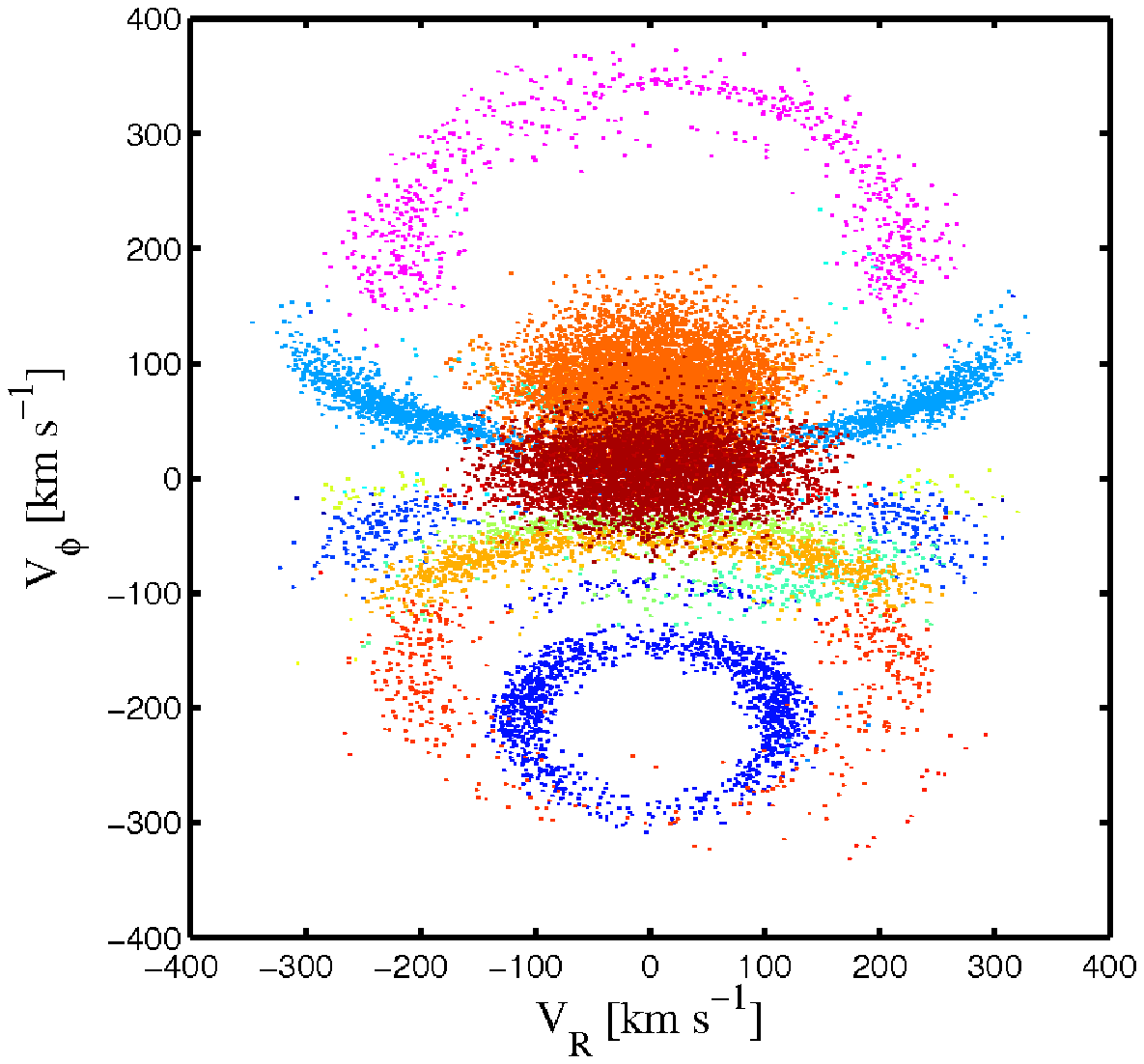}
\\
\includegraphics[width=80mm,clip]{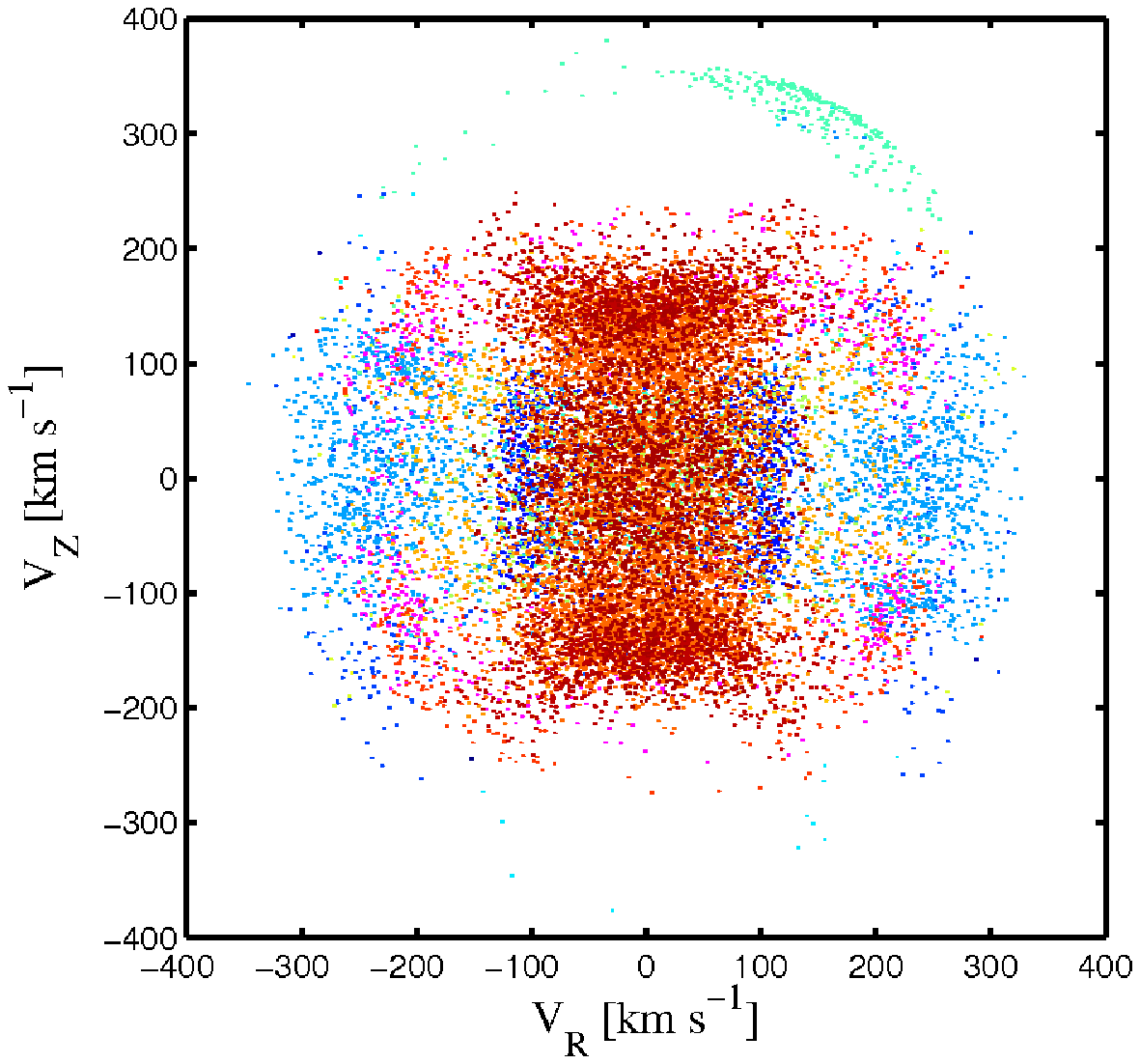}

\caption{Distribution in two different projections of velocity space
  of the stellar particles shown in Fig.~\ref{fig:mapas}.}
\label{fig:vel}
\end{centering}
\end{figure}

We  focus  now on  the  stellar particles  from  our  mock {\it  Gaia}
catalogue  (prior  to  error  convolution)  located  within  a  `Solar
Neighbourhood'  sphere of 2.5~kpc  radius, centred  at 8~kpc  from the
galactic centre.  Figures~\ref{fig:mapas} and \ref{fig:vel} show their
distribution on  the sky and in velocity  space, respectively.  Inside
this sphere we find  approximately $2 \times 10^{4}$ stellar particles
coming from 22 different satellites that contribute with, at least, 20
stars brighter than $M_{V} \approx  4.5$.  The distribution on the sky
is very  smooth and, thus,  disentangling merger debris in  the `Solar
Neighbourhood' by  only using spatial information is  not obvious.  On
the contrary, substructure can be  observed in velocity space but this
is  clearly less  sharply defined  than what  is found  in  the pseudo
integrals of  motion $E$--$L_{z}$ space  (e.g.  Figure~\ref{fig:ELz}).
We turn next to the space  of orbital frequencies where we also expect
to find much clumpiness.

\subsection{Frequency space}
\label{sec:sn_no_err}
The orbital frequencies of the stellar particles are computed as
follows (see also section~5.1 of GH10):
\begin{itemize}
\item{We assume a fixed underlying potential: as described in
  Section~\ref{pots} at $z=0$.}
\item{We integrate the orbits of each stellar particle for
  approximately 100 orbital periods.}
\item{We obtain their orbital frequencies by applying the spectral
  analysis code developed by \citet{daniel}.}
\end{itemize}

\subsubsection{Generalities}

The distribution of the stellar  particles in frequency space is shown
in the left panel of Figure~\ref{fig:omegas_n_error}.  We use the same
colour coding as in previous  figures. Note that satellite galaxies do
not  appear  as  a  single  and  smooth structure,  but  rather  as  a
collection of well  defined and small clumps.  These  small clumps are
associated  to  the  different  stellar  streams  crossing  our  Solar
Neighbourhood  sphere   at  present  time.   In   a  time  independent
gravitational  potential, streams are  distributed in  frequency space
along lines of constant $\Omega_{r}$ and $\Omega_{\phi}$.  However, in
a time dependent  potential, such as the one  considered here, streams
are distributed along lines with a given curvature, depending upon the
rate of growth of the potential (see also GH10).

The left-hand panel of Figure~\ref{fig:omegas_n_error} also shows that
in  the  $\Omega_{\phi}$   vs.   $\Omega_{r}  -  \Omega_{\phi}$  space
satellites may  strongly overlap.  This situation  can be considerably
improved by  adding the  $z$-component of the  angular momentum  as an
extra dimension  to this  space, as shown  in the right-hand  panel of
Figure~\ref{fig:omegas_n_error}.

The   distribution  of   debris  in   the  space   of   $\Omega_{r}  -
\Omega_{\phi}$  vs.   $L_{z}$  is  very  comparable  to  that  in  $E$
vs. $L_{z}$,  as a coarse comparison  between Figure~\ref{fig:ELz} and
Figure~\ref{fig:omegas_n_error} will reveal.  Note  as well that as in
$E-L_{z}$ space,  from this projection  it is possible, solely  by eye
inspection, to isolate a few accreted satellites.

From this Figure  we can also notice that  satellites on low frequency
orbits have a smaller amount and  a set of better defined streams than
those on orbits with high  frequency.  The reason for this is twofold.
Firstly, satellites on  low frequency orbits spend most  of their time
far from  the centre of the  host potential and  therefore have longer
mixing  time-scales, as  opposed  to those  on  high frequency  orbits
(short periods).   Secondly, potentials as the one  considered in this
work may admit a  certain degree of chaoticity \citep[e.g.][]{schall}.
Satellites on highly  eccentric short period orbits come  close to the
galactic centre and may be `scattered' via chaos.  Such chaotic orbits
do  not have  stable fundamental  frequencies (since  they  have fewer
integrals of motion than  needed and thus wander through phase-space).
Therefore, their structure in frequency space is rapidly erased.

It  is  important  to  note  that  our  desire  to  analyse  a  `Solar
Neighbourhood'  sphere  has   resulted  in  certain  limitations.  For
example,  some satellites  do not  contribute  at all  to this  volume
because  of their  particular orbit.  In addition,  most of  the faint
satellite  galaxies have  only a  few, if  any, `stars'  with absolute
magnitude  above  the  limiting  magnitude  of  our  Mock  {\it  Gaia}
catalogue and therefore they are not `observable', either.

\subsubsection{Estimating the time of accretion}
\label{sec:time_acc}

\begin{figure*}
\centering
\includegraphics[width=80mm,clip]{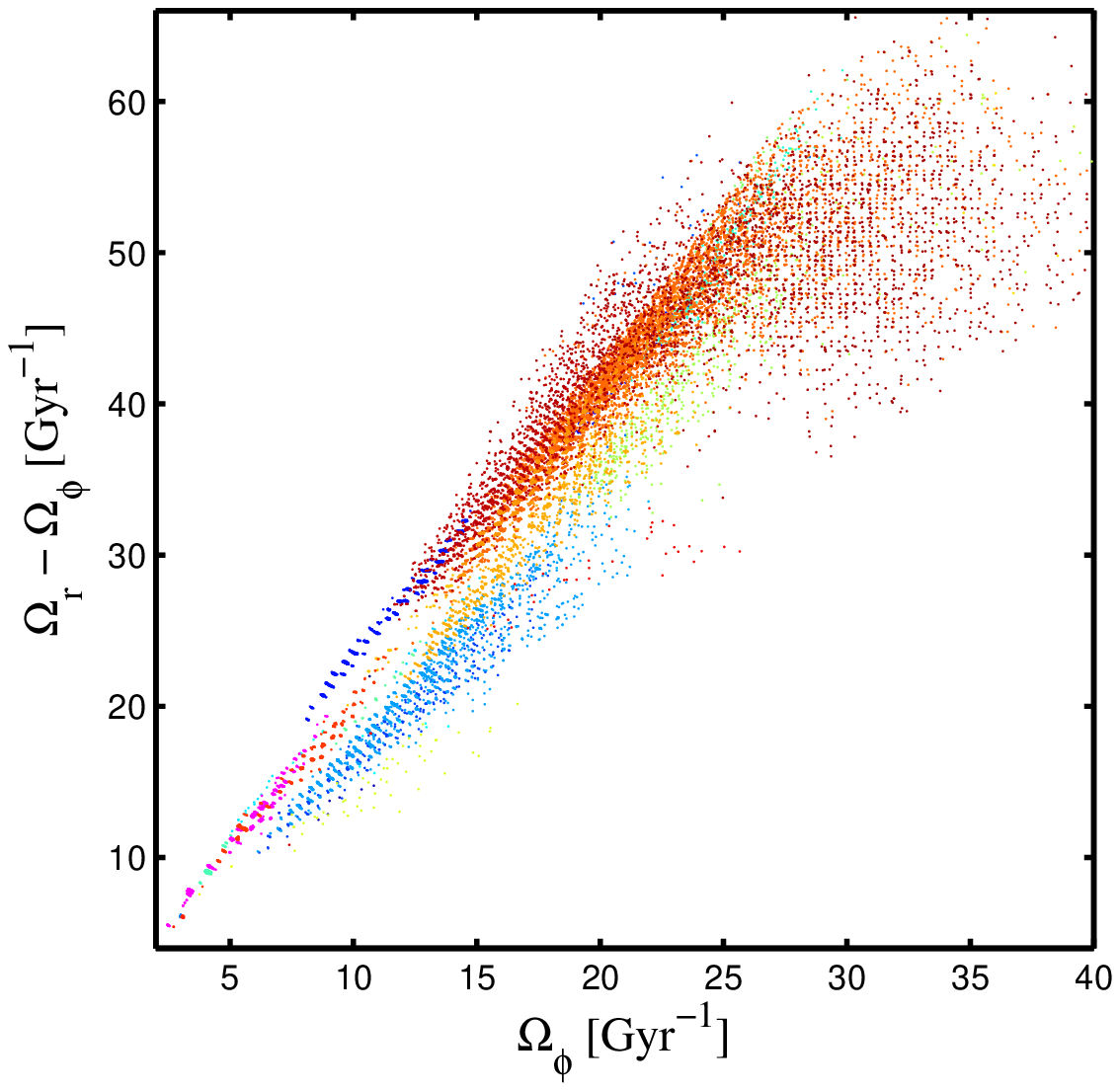}
\hspace{0.001cm}
\includegraphics[width=80mm,clip]{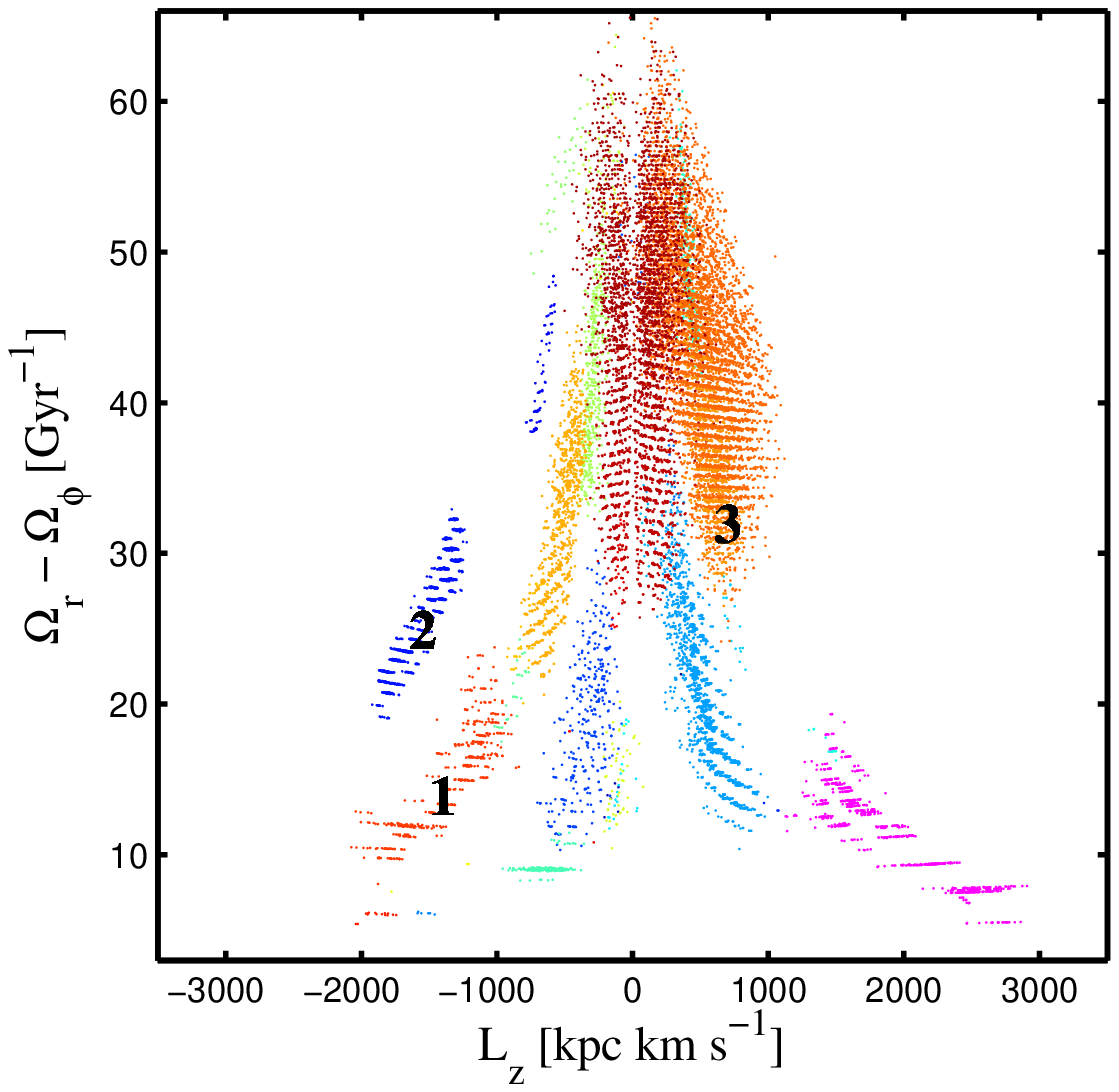}
\caption{Distribution  of stellar  particles  in frequency  (left-hand
  panel)  and $L_{z}$  vs.  $\Omega_{r}  -  \Omega_{\phi}$ (right-hand
  panel) space located inside a sphere of 2.5~kpc radius at 8~kpc from
  the galactic centre.   The colour-coding is the same  as in previous
  figures.  Note  that this  distribution  of  particles was  obtained
  without taking  into account  the expected {\it  Gaia} observational
  errors.}
\label{fig:omegas_n_error}
\end{figure*}

An important  feature of  frequency space is  that an estimate  of the
time since  accretion of a satellite  can be derived  by measuring the
separation  between  adjacent  streams  along the  $\Omega_{\phi}$  or
$\Omega_{r}$  directions   \citep[][GH10]{mcm}.   This  characteristic
scale may be estimated through a Fourier analysis technique as follows

\begin{itemize} 
\item{We create  an image of the  scatter plot in  frequency space, by
  griding  the space  with a  regular $N  \times N$  mesh of  bin size
  $\Delta$ and count the number of stellar particles on each bin.}
\item{We  apply a  2-D discrete  Fourier Transform  to this  image and
  obtain $H(k_{r},k_{\phi})$}
\item{We compute  a one-dimensional power  spectrum along a  thin slit
  centred on the wavenumbers $k_{r}/(N\Delta),~k_{\phi}/(N\Delta) = 0$
  as
\begin{equation}
\begin{array}{llll}
\displaystyle
\label{ps_estimation}
P(0)=\frac{1}{N^2} \left|H(0,0)\right|^2, \\
\\
\displaystyle
P(k_{\phi})=\frac{1}{N^2}\left[\left|H(k_{\phi},0)\right|^2+\left|H(-k_{\phi},0)\right|^2\right]\\
\\
$for $ k_{\phi}=1,\dots,\left(\frac{N}{2}-1\right),\\
\\
\displaystyle
P(N/2)=\frac{1}{N^2} \left|H(-N/2,0)\right|^2 \\
\end{array}
\end{equation}
and analogously for $P(k_{r})$.}
\item{We identify the  wavenumber $f_{0}$ of the dominant  peak in the
  spectra,  which corresponds to  a wavelength  equal to  the distance
  between adjacent streams in frequency space.}
\item{  The estimate of  the time  since accretion  is $\tilde{t}_{\rm
    acc} = 2\pi f_{0}$ (for more details, see section~3.2.3 of GH10)}.

\end{itemize}

 We have  applied this method  to three different satellites  from our
 simulations. Two of these satellites  (labelled number 1 and 2 in the
 right-hand  panel  of  Figure~\ref{fig:omegas_n_error})  have  a  low
 frequency guiding orbit and, just  by eye inspection, can be isolated
 in  $L_{z}$ vs.   $\Omega_r -  \Omega_{\phi}$ space.   The  third one
 (number 3) has a high  frequency guiding orbit and overlaps with some
 other satellites in this space.

\begin{figure*}

\includegraphics[width=61.3mm,clip]{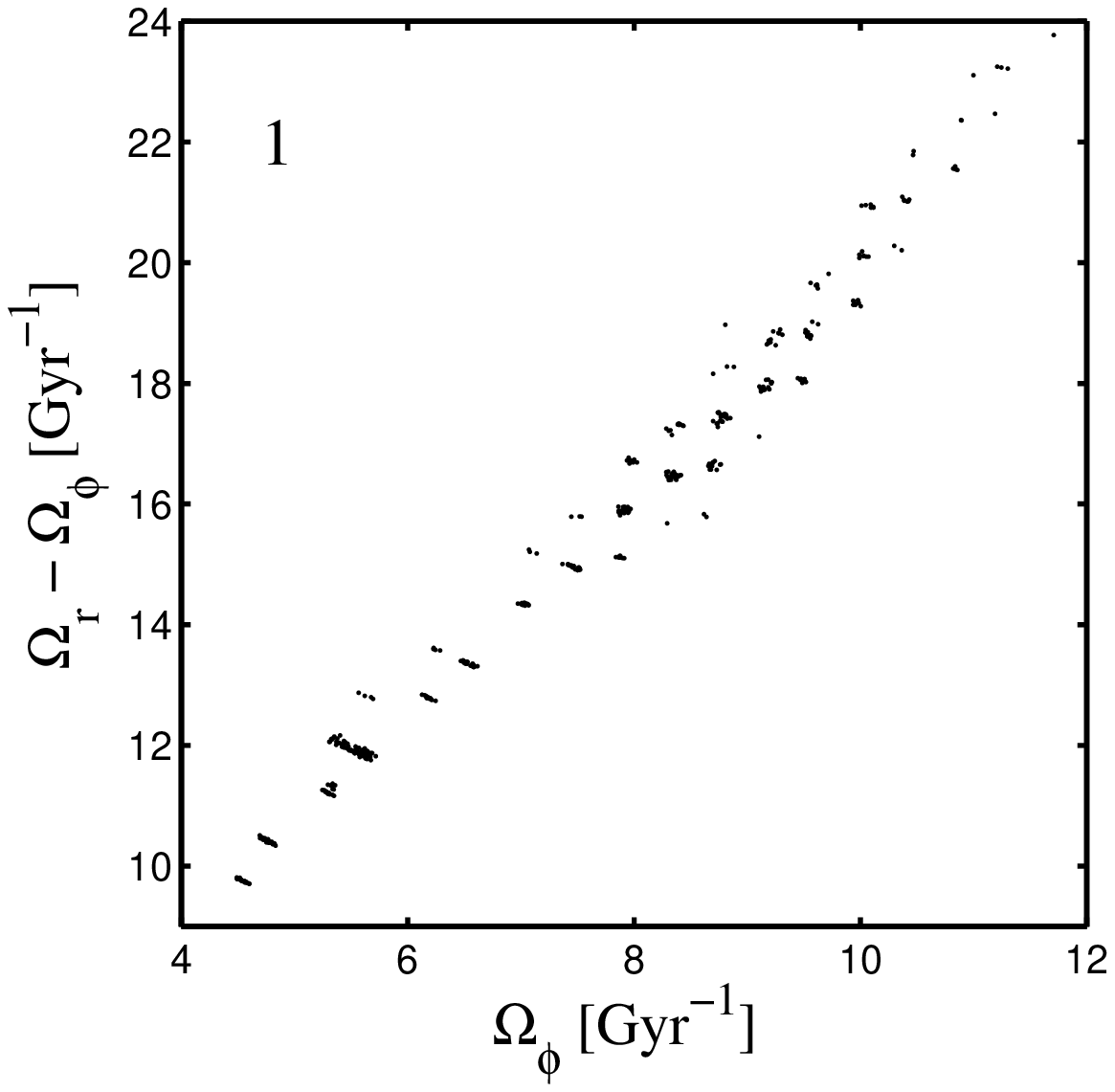}
\hspace{0.001cm}
\includegraphics[width=55mm,clip]{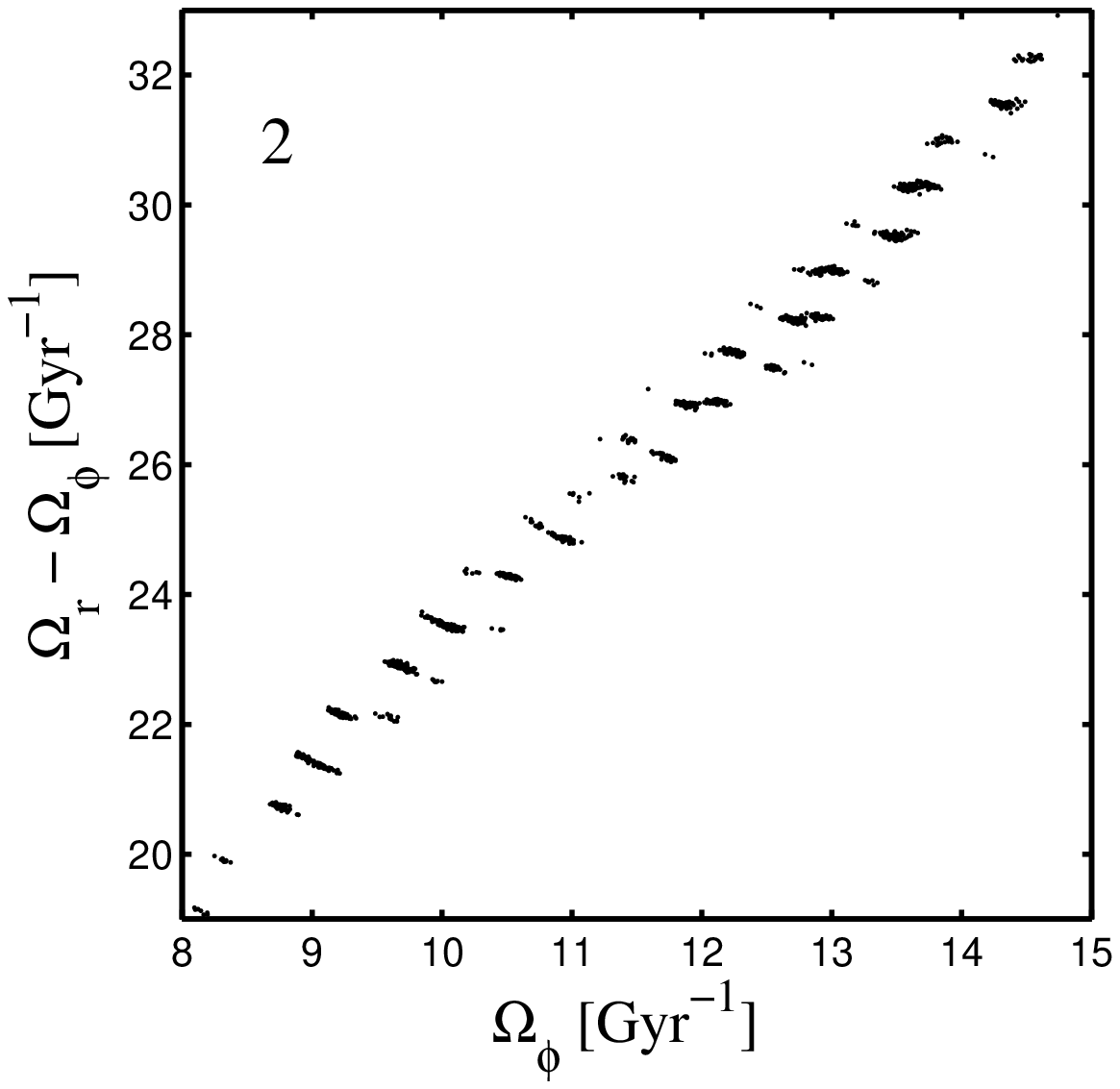}
\hspace{0.001cm}
\includegraphics[width=55mm,clip]{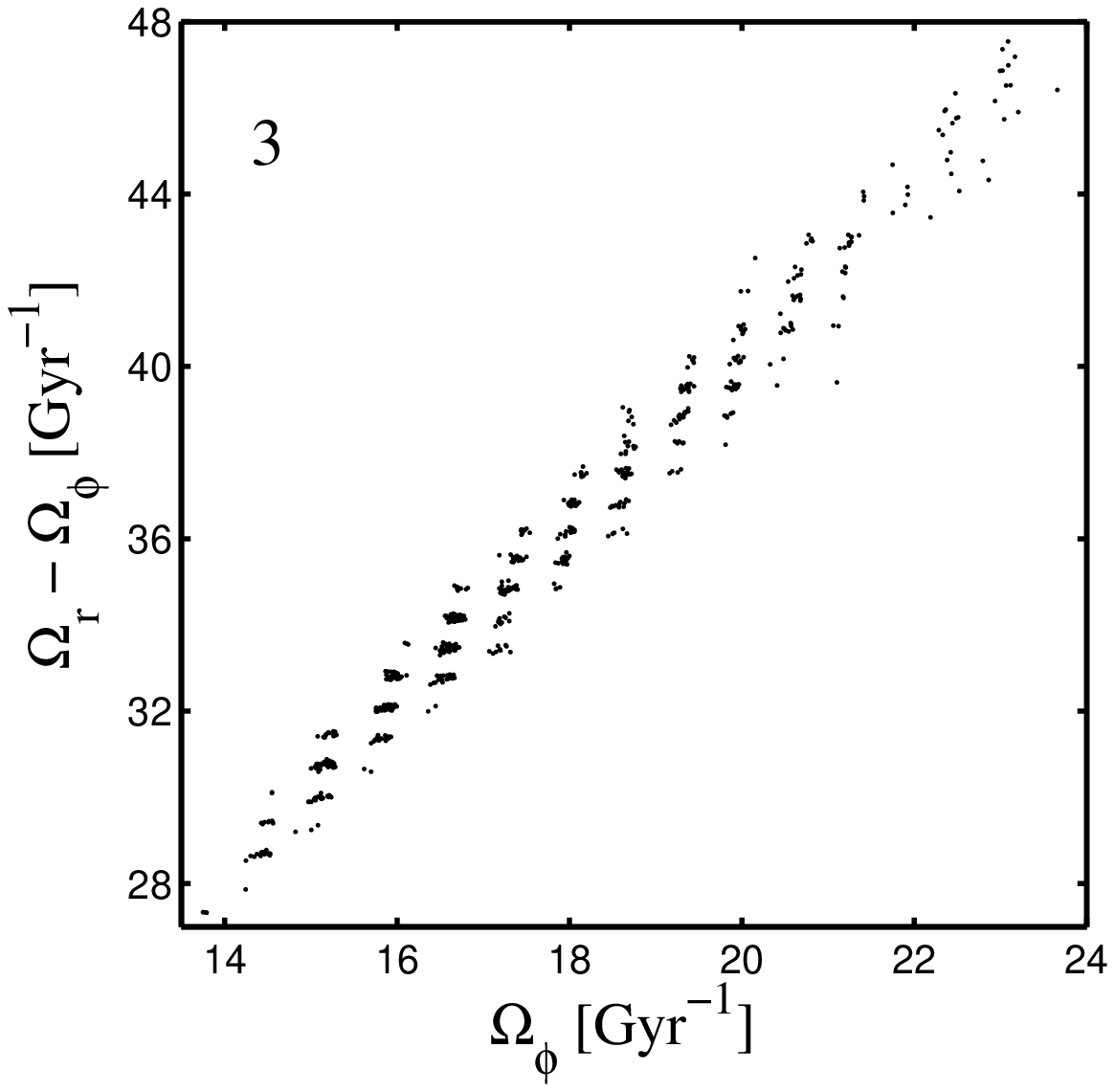}
\\
\includegraphics[width=58mm,clip]{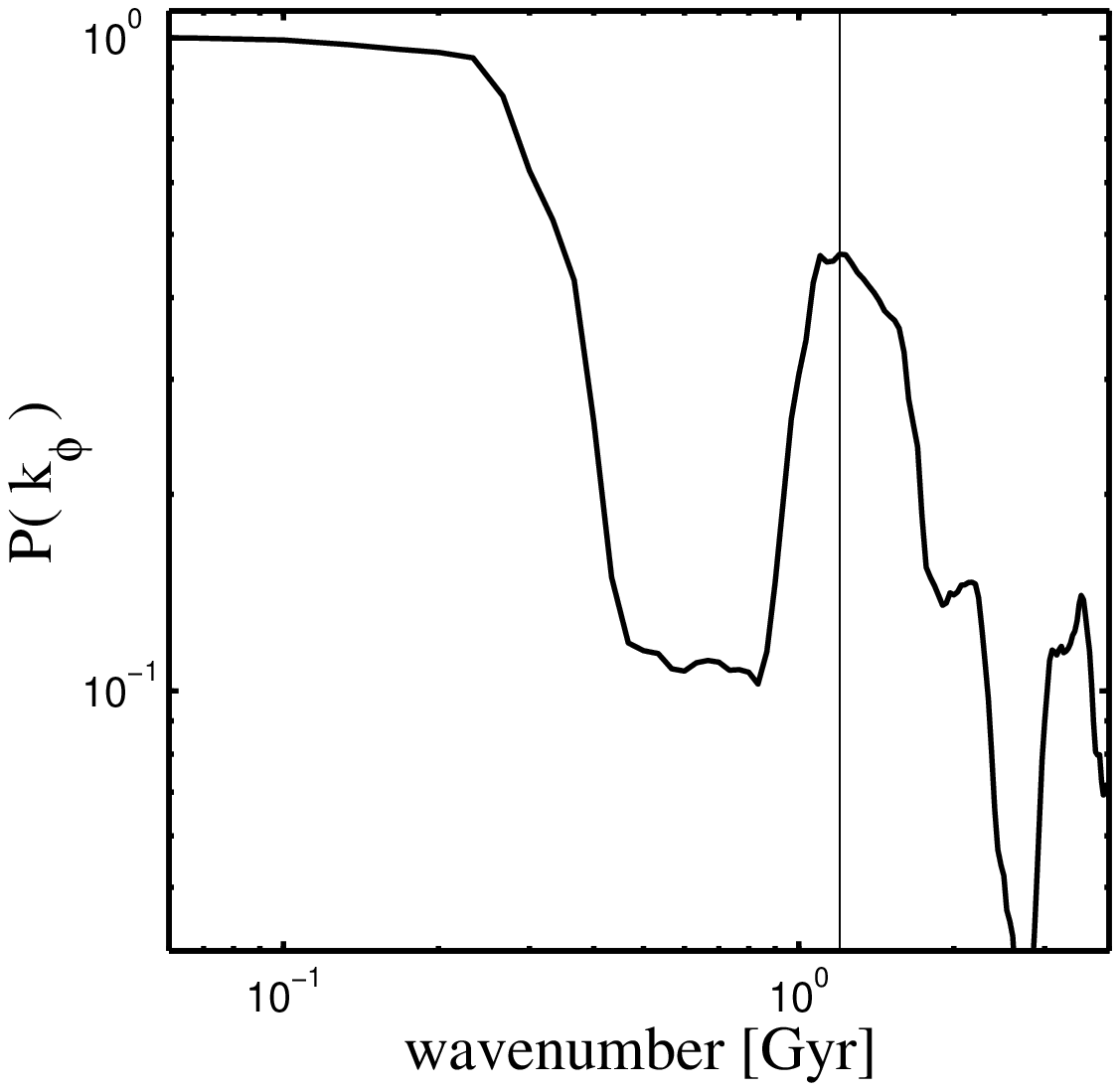}
\hspace{0.001cm}
\includegraphics[width=55mm,clip]{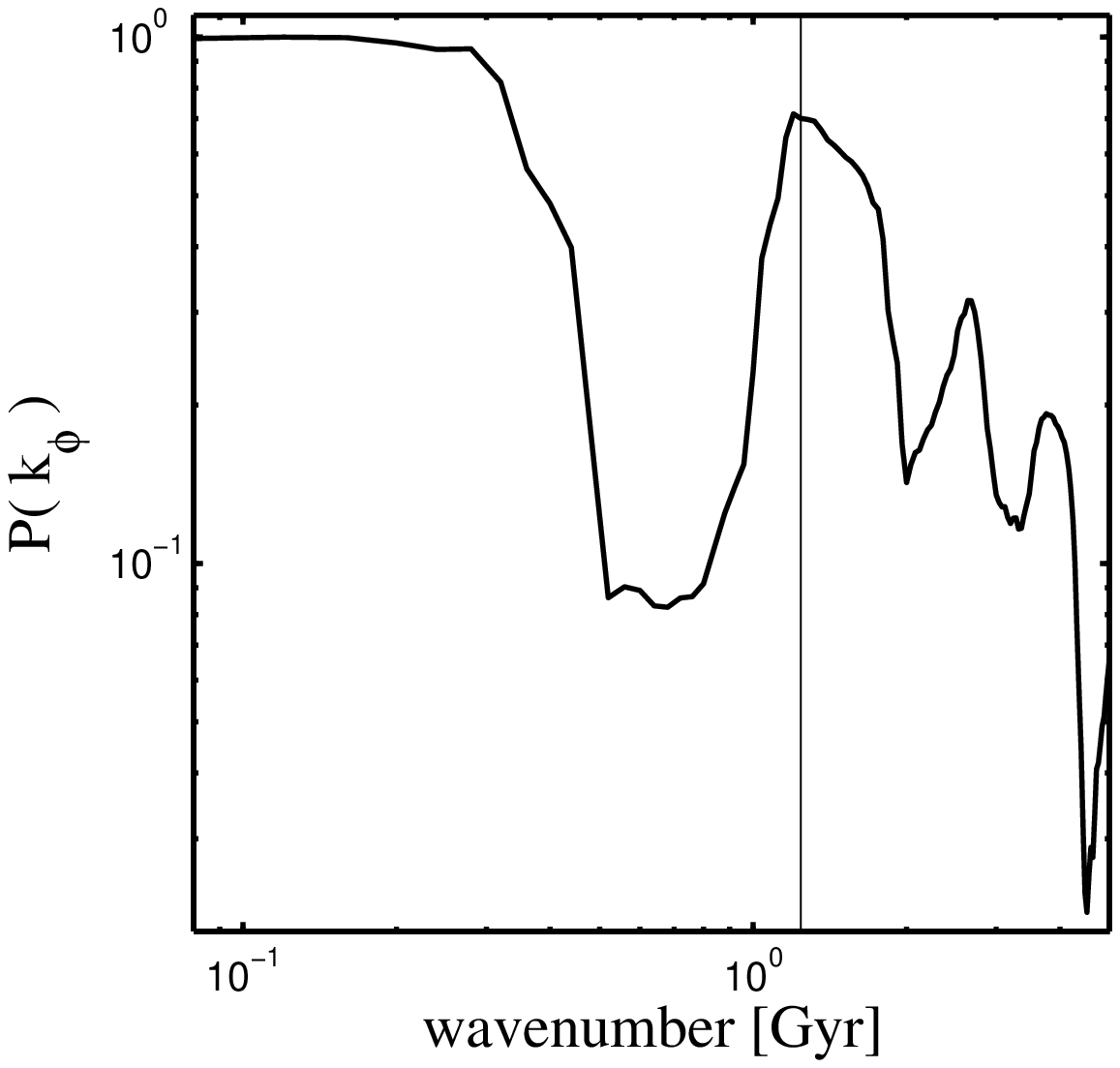}
\hspace{0.001cm}
\includegraphics[width=55mm,clip]{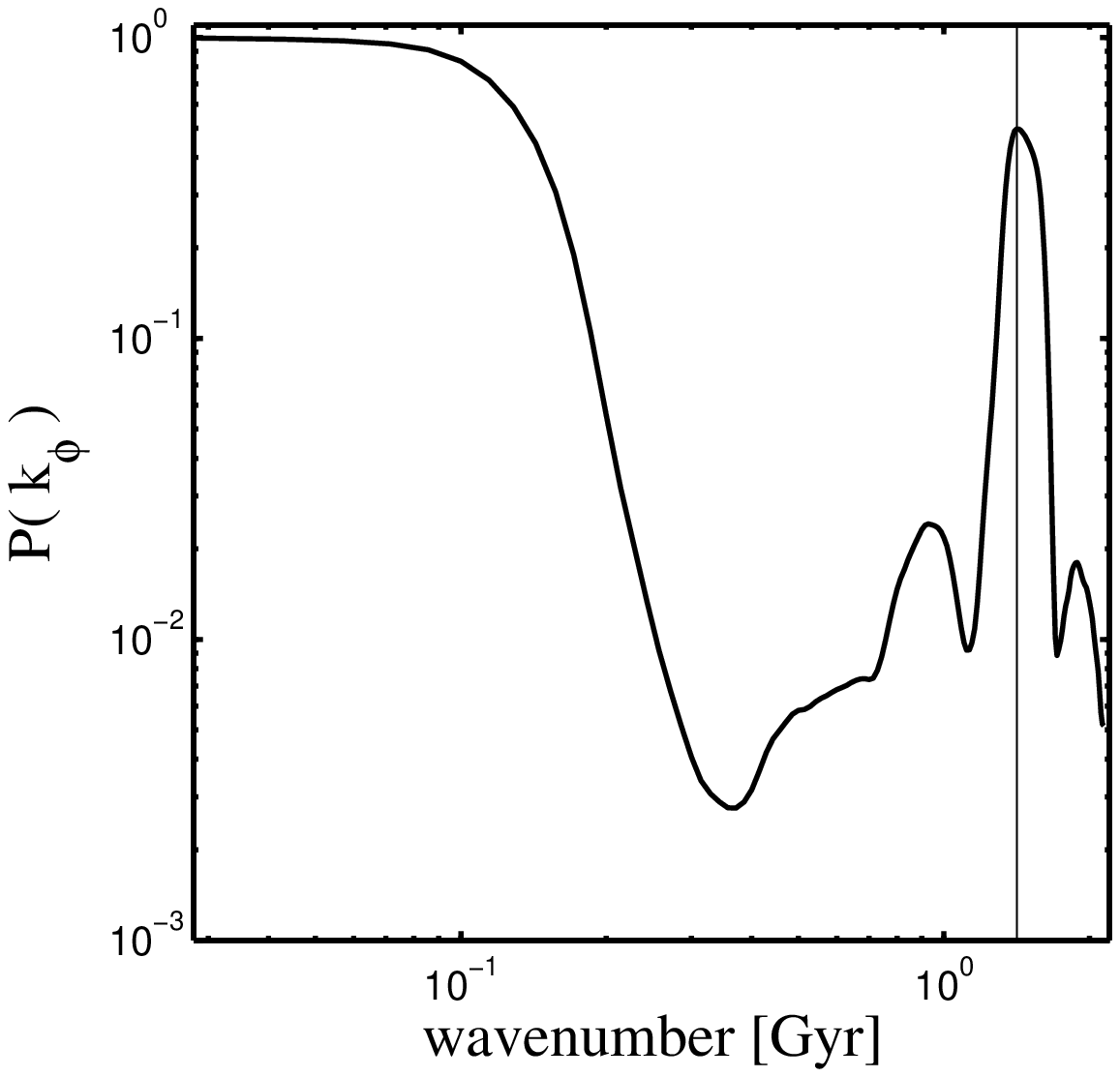}

\caption{The  top  panels show  for  three  different satellites,  the
  distribution of stellar particles  in frequency space located inside
  a sphere of  2.5~kpc radius at 8~kpc from  the galactic centre.  The
  bottom  panels  show the  1-D  normalized  power  spectra along  the
  $k_{r}=0$      direction,     obtained      as      explained     in
  Section~\ref{sec:time_acc}.   The wavenumber  of  the dominant  peak
  (denoted by the vertical lines  in these panels) is used to estimate
  the accretion time of the satellites. Other peaks in the spectra are
  associated  to either  the harmonics  of this  wavenumber or  to the
  global shape of the particle's distribution in frequency space.  }
\label{fig:sats_frq}
\end{figure*}

In   Figure~\ref{fig:sats_frq}  we   zoom-into  the   distribution  of
particles in frequency space for each satellite separately.  From this
figure we clearly appreciate the  large number of streams that each of
these satellites is contributing to this `Solar Neighbourhood' volume.
Although the number  of streams apparent in this  Figure is very large
(44,  30 and  59, for  satellite 1,  2 and  3, respectively),  this is
consistent with  the models  of \citet{hw}, who  predicted a  total of
$\sim 300  - 500$  streams locally, i.e.   in an  infinitesimal volume
around the  Sun.  This is encouraging particularly  because the models
presented here are much more realistic than those by \citet{hw}.

To obtain an  estimate of the time since  accretion for our satellites
we  compute  the normalized  power  spectrum  along  the $k_{r}  =  0$
direction.    The  results  are   shown  in   the  bottom   panels  of
Figure~\ref{fig:sats_frq}.   From left  to  right, the  peak with  the
highest amplitude in the spectrum is located at a wavenumber of $f_{0}
= 1.24$, $1.41$ and  $1.21$~Gyr, respectively. These values correspond
to an  estimated time since  accretion of $7.8$, $8.9$  and $7.6$~Gyr,
respectively. These  are in reasonable agreement with  the true value,
which  we define  as the  time when  $80\%$ of  the  stellar particles
became  unbound\footnote{For  simplicity  we  assume that  a  particle
becomes unbound when it is found outside the initial satellite's tidal
radius, obtained from the  initial King profile.}.  For each satellite
we obtain a  $t_{\rm acc}$ of $9$, $9.4$  and $9.7$~Gyr, respectively.
In congruence  with GH10,  we find that  this method provides  a lower
limit to  the actual time since  accretion, differing by  $15 - 25\%$.
This  is very  encouraging since  GH10 did  not study  such  a complex
growth of the host potential.

\section{Analysis of the mock {\it Gaia} catalogue}

We will now analyse  the phase-space distribution of stellar particles
located in a  volume in the `Solar Neighbourhood',  as may be observed
in the future  by {\it Gaia}.  We study how the  {\it Gaia} errors, as
well as the contamination from other Galactic components, would affect
our ability to identify and characterize merger debris.

\subsection{Contamination by disc and bulge}
As  in Section~\ref{sec:sn_no_err},  we restrict  our analysis  to the
stellar particles located inside  a sphere of $2.5$~kpc radius centred
at  $8$~kpc from  the galactic  centre.   To account  for the  maximum
possible degree of background contamination, we consider all the stars
for which {\it Gaia}  will measure full phase-space coordinates (i.e.,
all stars brighter than $V=17$), according to the Monte Carlo model of
the disc and bulge.

Disc particles largely outnumber  those in the stellar halo, generally
providing a prominent background. However  much of this can be removed
by  considering the  Metallicity  Distribution Function  (MDF) of  the
Galactic components.  While the MDF of the halo peaks at approximately
$[{\rm Fe}/{\rm H}] = -1.6$, that of disc peaks at $[{\rm Fe}/{\rm H}]
\approx -0.2$.  As  a consequence, stars from the  disc are in general
more metal-rich and therefore a  simple cut on metallicity could be of
great help to isolate halo stars in our sample.

It is therefore important to characterize well the metal poor tail of
the Galactic disc MDF. We use the model by \citet{ive08} to represent
this tail:
\begin{equation}
p(x) = G_{1}(x|\mu_{1},\sigma_{1}) + 0.2 ~ G_{2}(x|\mu_{2},\sigma_{2}), \qquad
x=[{\rm Fe}/{\rm H}],
\end{equation}
and
\begin{equation}
G(x|\mu,\sigma) = \frac{1}{\sqrt{2\pi}\sigma} \exp{\frac{-(x-\mu)^{2}}{2\sigma^{2}}}.
\end{equation}
Here $\sigma_{1} = 0.16~\text{dex}$, $\sigma_{2} = 0.1~\text{dex}$,
$\mu_{2} = -1.0$. \citet{ive08} found $\mu_{1}$ to vary as a function
of the height from the plane of the disc, but for simplicity we fix
$\mu_{1} = -0.71$. The contribution of stars more metal-rich than
$[{\rm Fe}/{\rm H}] \approx -0.4$ to the study of \citet{ive08} is
small (because of a restriction imposed on the colour range). Therefore
to account for this population in our model, we normalize $p([{\rm
    Fe}/{\rm H}])$ with a constant $\alpha$. The numerical value of
$\alpha$ is such that at its peak value (located at $[{\rm Fe}/{\rm
    H}] \approx -0.7$) $p([{\rm Fe}/{\rm H}])$ matches the
MDF of the Geneva-Copenhagen survey \citep{gc}.

\begin{figure*}
\includegraphics[width=80mm,clip]{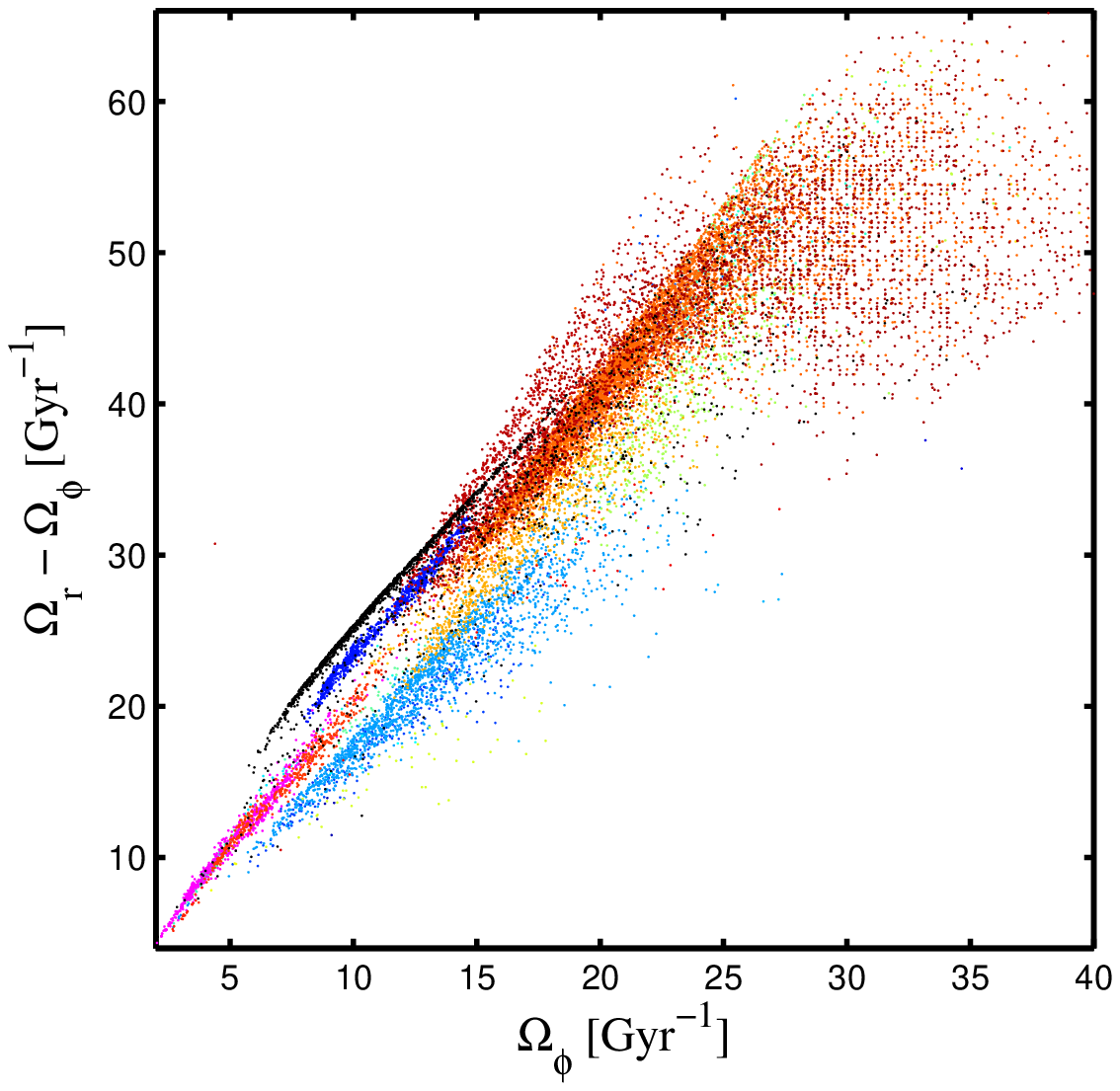}
\hspace{0.001cm}
\includegraphics[width=80mm,clip]{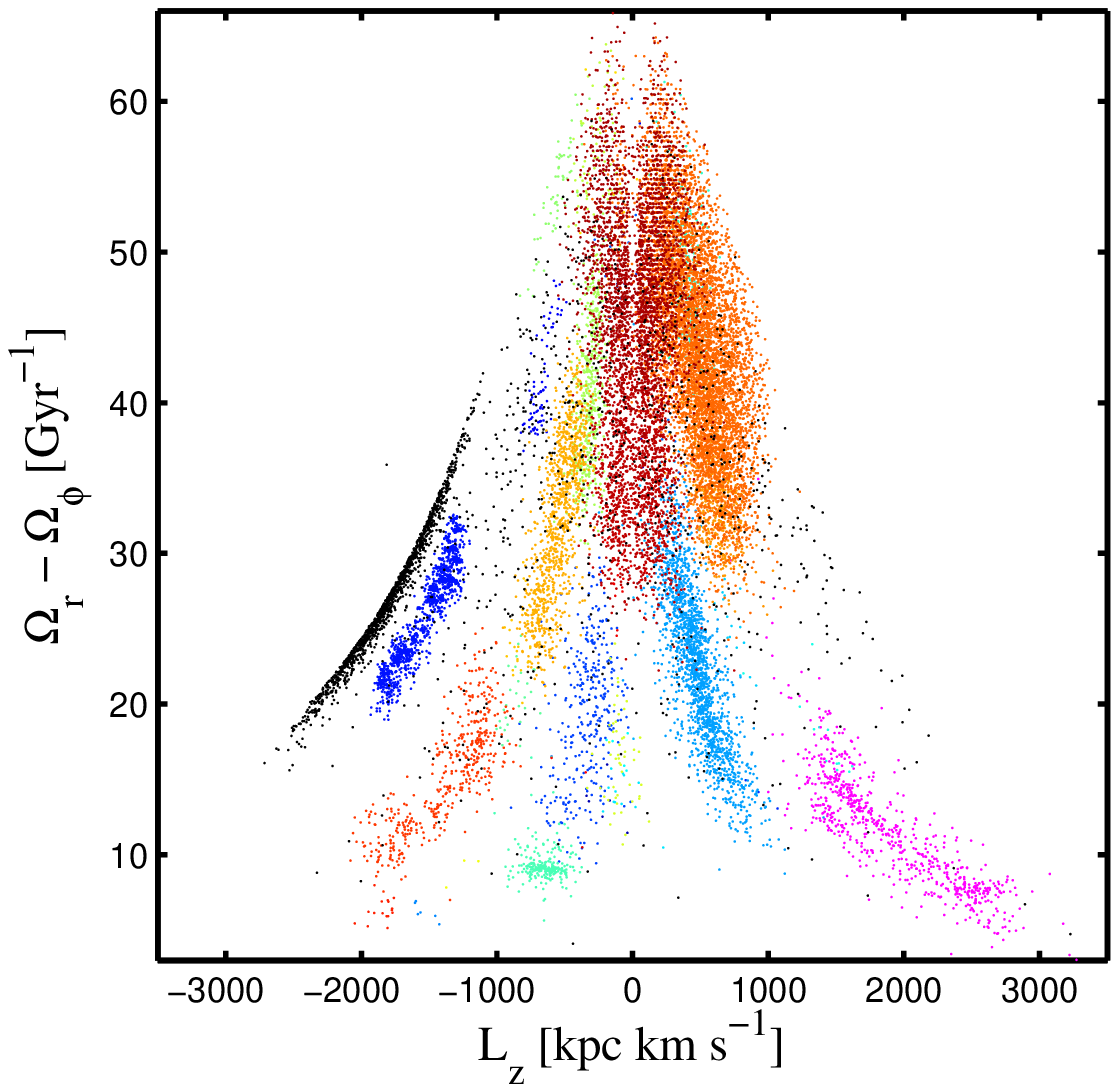}
\caption{Distribution of stellar particles in frequency (left-hand panel)
  and $L_{z}$ vs. $\Omega_{r} - \Omega_{\phi}$ (right-hand panel) space
  located inside a sphere of 2.5~kpc radius at 8~kpc from the galactic
  centre, after convolution with the {\it Gaia} observational
  errors. The black dots denote the contribution from the disc and
  bulge. The rest of the colours represent different satellites.}
\label{fig:w_err}
\end{figure*}

Having derived a MDF for our disc model, we can proceed to apply a cut
on metallicities $[{\rm Fe}/{\rm H}]  \geq -1.1$ to the disc particles
present in our `Solar Neighbourhood'  sphere of 2.5~kpc radius.  Although
in  our simulation halo  stars have  metallicities between  $-2.0 \leq
[{\rm  Fe}/{\rm H}] \leq  -1.5$, stars  with higher  metallicities are
known  to be  present in  the Galactic  stellar halo.  Hence  the more
generous  cut. This  cut leads  to a  subsample of  only  $\approx 9.3
\times 10^{3}$ `disc stars' brighter than $V=17$ out  of a total of
$4.1 \times 10^{7}$ found in our model.

A similar  approach can be followed  for the bulge  component. In this
case metallicities are  assigned according to the observed  MDF of the
Galactic bulge,  as given  by \citet{zoca}. This  MDF is modelled  as a
single  gaussian distribution  with a  mean  metallicity $\langle[{\rm
    Fe}/{\rm H}]\rangle = -0.28$ {\rm  dex} and a dispersion $\sigma =
0.4$  {\rm  dex},  as  found   in  the  outermost  field  analysed  by
\citet{zoca} (located at a latitude of $b = -12^{\circ}$).

 After removing  all stellar particles  with $[{\rm Fe}/{\rm  H}] \geq
 -1.1$, out of the $2.9 \times 10^{4}$ bulge stellar particles with $V
 \le 17$  originally present in  our `Solar Neighbourhood' sphere  we are
 left with  only 784 stars. This  very small number  suggests that the
 contamination from this galactic component should not strongly affect
 the detection of substructure in phase-space.

\subsection{Frequency Space}

Figure~\ref{fig:w_err}  shows the  distribution  of stellar  particles
inside the  `Solar Neighbourhood' sphere in  both frequency (left-hand
panel)  and  $L_{z}$  vs.   $\Omega_{r} -  \Omega_{\phi}$  (right-hand
panel)  space,  now after  error  convolution.   Again, the  different
colours indicate stellar particles from different satellites, with the
addition   of    the   disc   and    bulge   which   are    shown   in
black. Interestingly, in both spaces disc particles are part of a very
well defined and quite isolated structure since, as expected for stars
moving  in the  galactic  plane on  a  circular orbit,  they have  the
largest values  of $|L_{z}|$ at a given  $\Omega_{r} - \Omega_{\phi}$.
Therefore,  it  is  possible  to  isolate  and  easily  eliminate  the
contamination from  the disc. On  the other hand, bulge  particles are
sparsely distributed in both spaces  and, although there are very few,
they  can  not  be  simply  isolated  as in  the  case  of  the  disc.
Nonetheless, they  do not define a  clump that could  be confused with
merger debris.
 
Both   panels  of   Figure~\ref{fig:w_err}  show   that,   even  after
convolution  with  the latest  model  of  the  {\it Gaia}  measurement
errors,   significant  substructure   is  apparent   in   $L_{z}$  vs.
$\Omega_{r} -  \Omega_{\phi}$ vs.  $\Omega_{\phi}$  space.  Comparison
to  Figure~\ref{fig:omegas_n_error}  shows  very  good  correspondence
between  clumps,   although  these  are,  as  a   consequence  of  the
convolution,  generally less  well defined  and contain  less internal
substructure.

\subsubsection{Estimating the time since accretion}
\label{sec:est_time_acc_err}

\begin{figure*}
\hspace{0.001cm}
\includegraphics[width=61.5mm,clip]{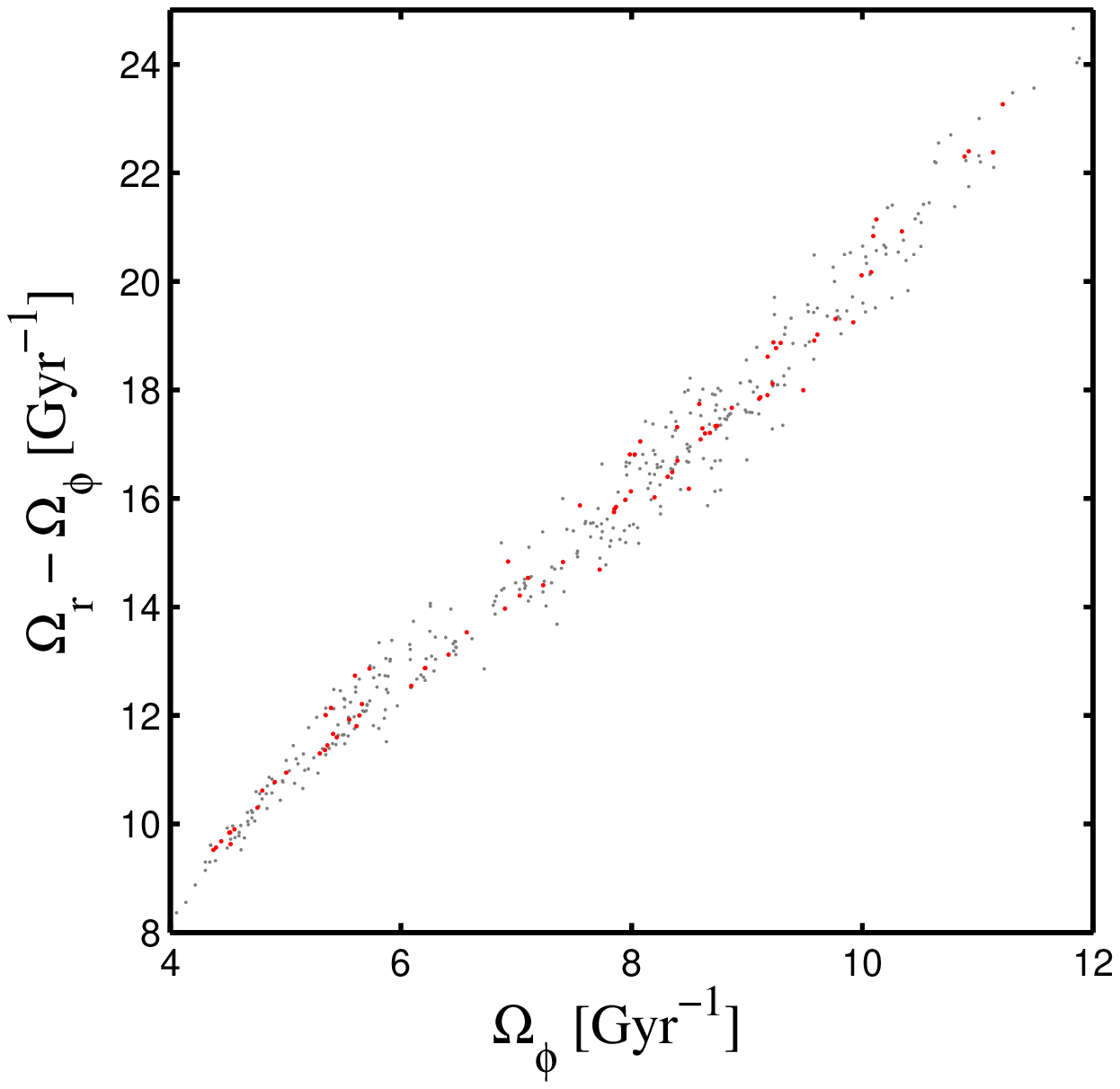}
\hspace{0.001cm}
\includegraphics[width=55mm,clip]{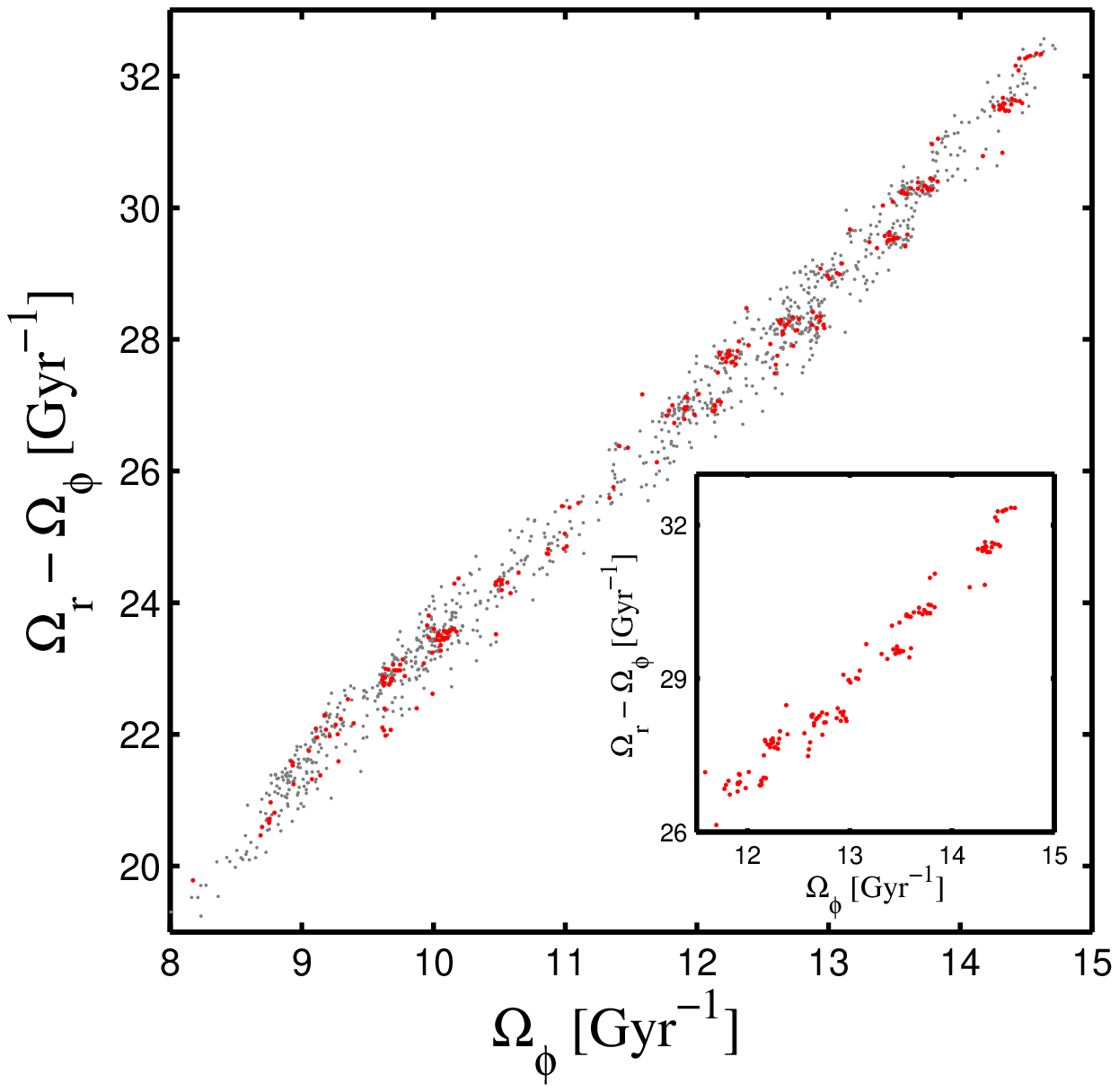}
\hspace{0.001cm}
\includegraphics[width=55mm,clip]{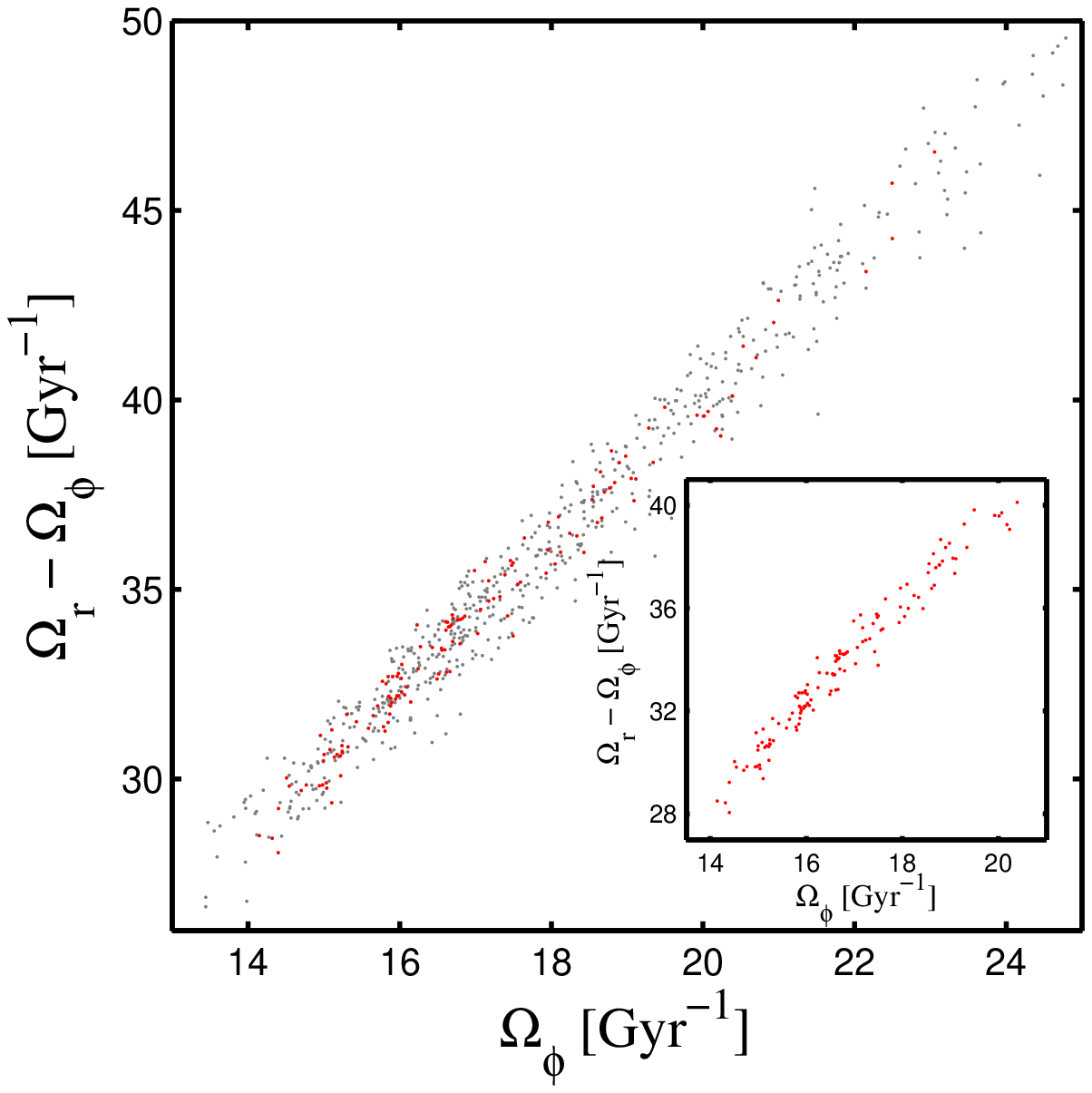}
\\
\includegraphics[width=60.5mm,clip]{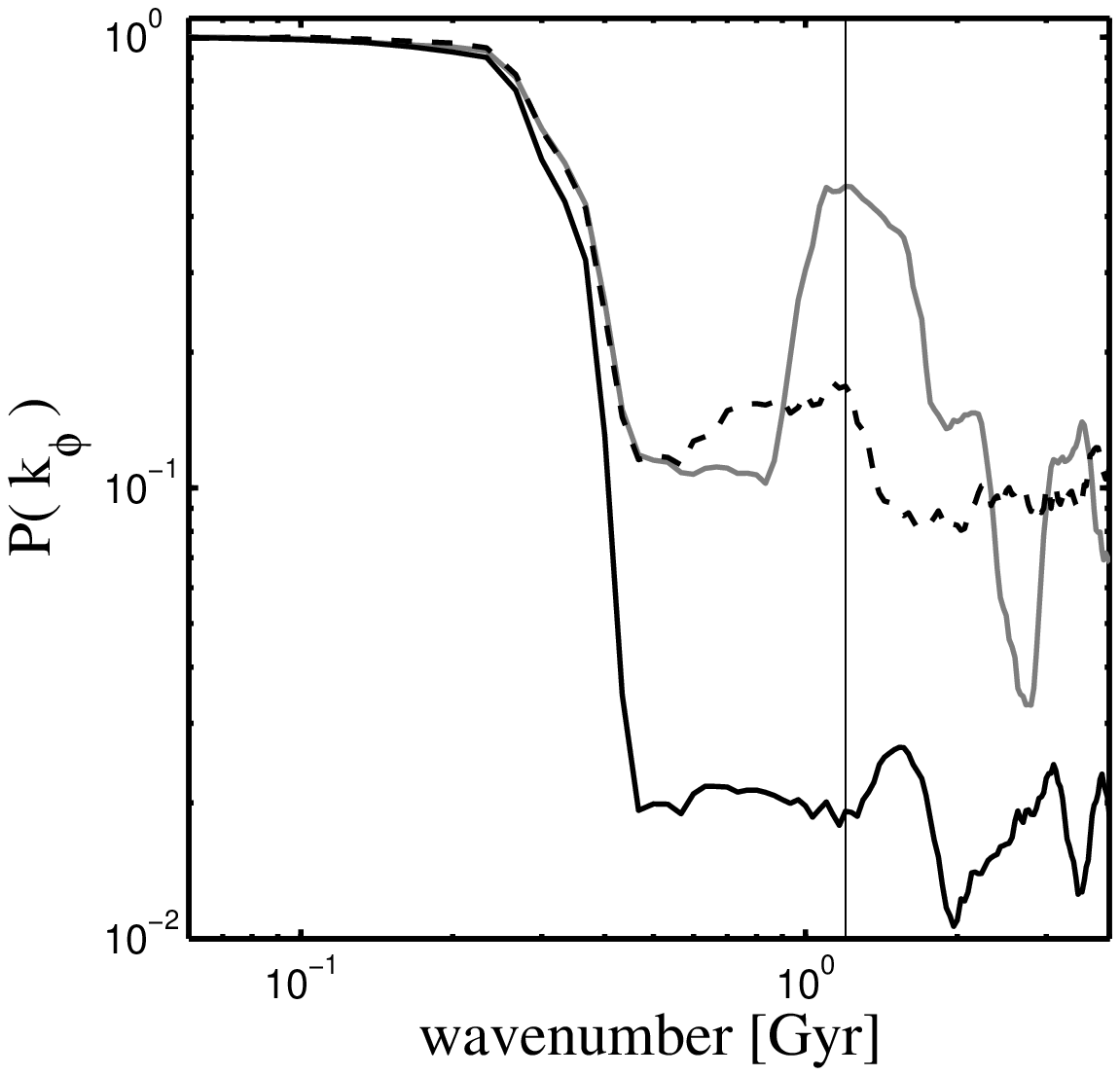}
\hspace{0.001cm}
\includegraphics[width=55mm,clip]{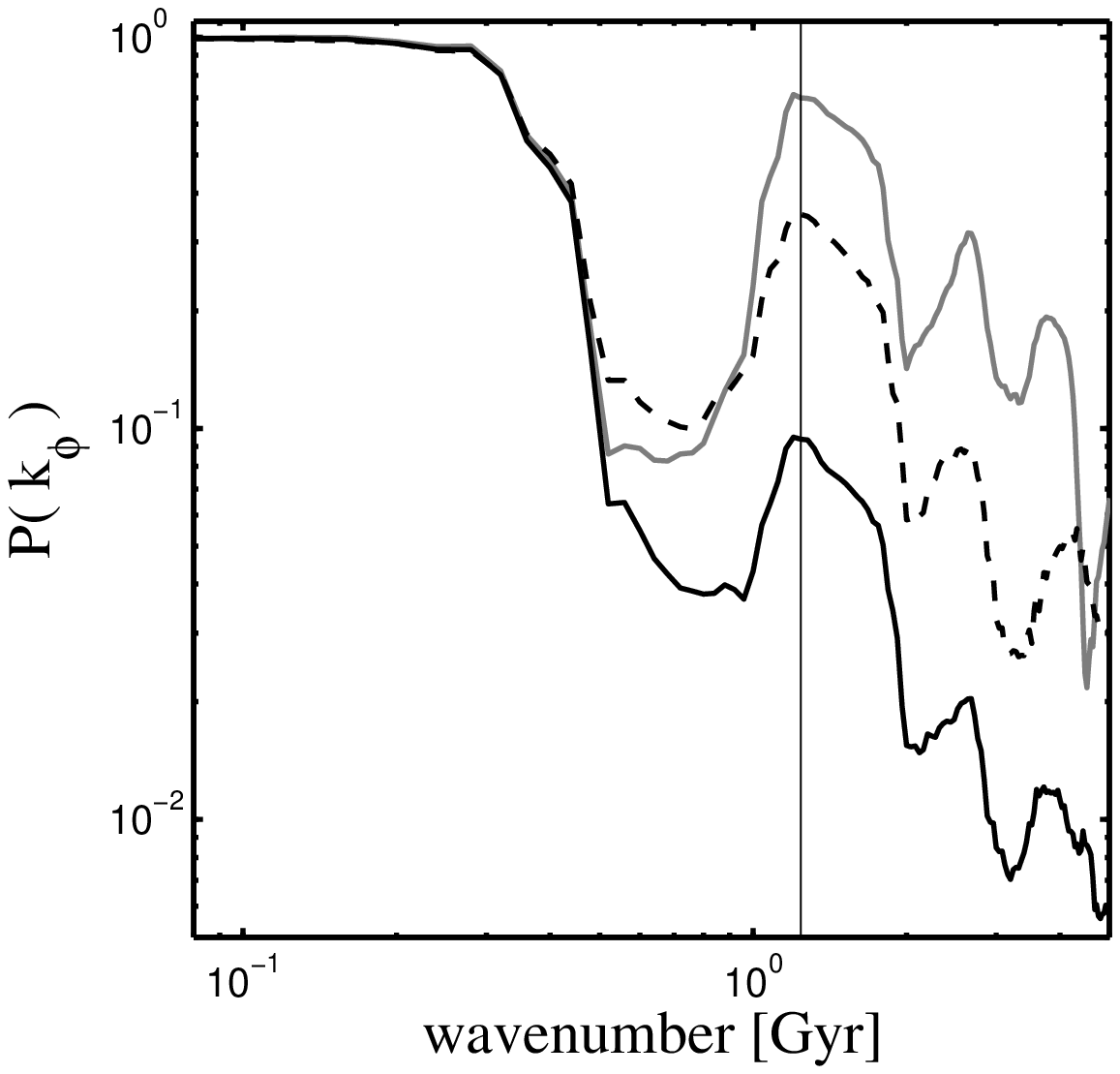}
\hspace{0.001cm}
\includegraphics[width=55mm,clip]{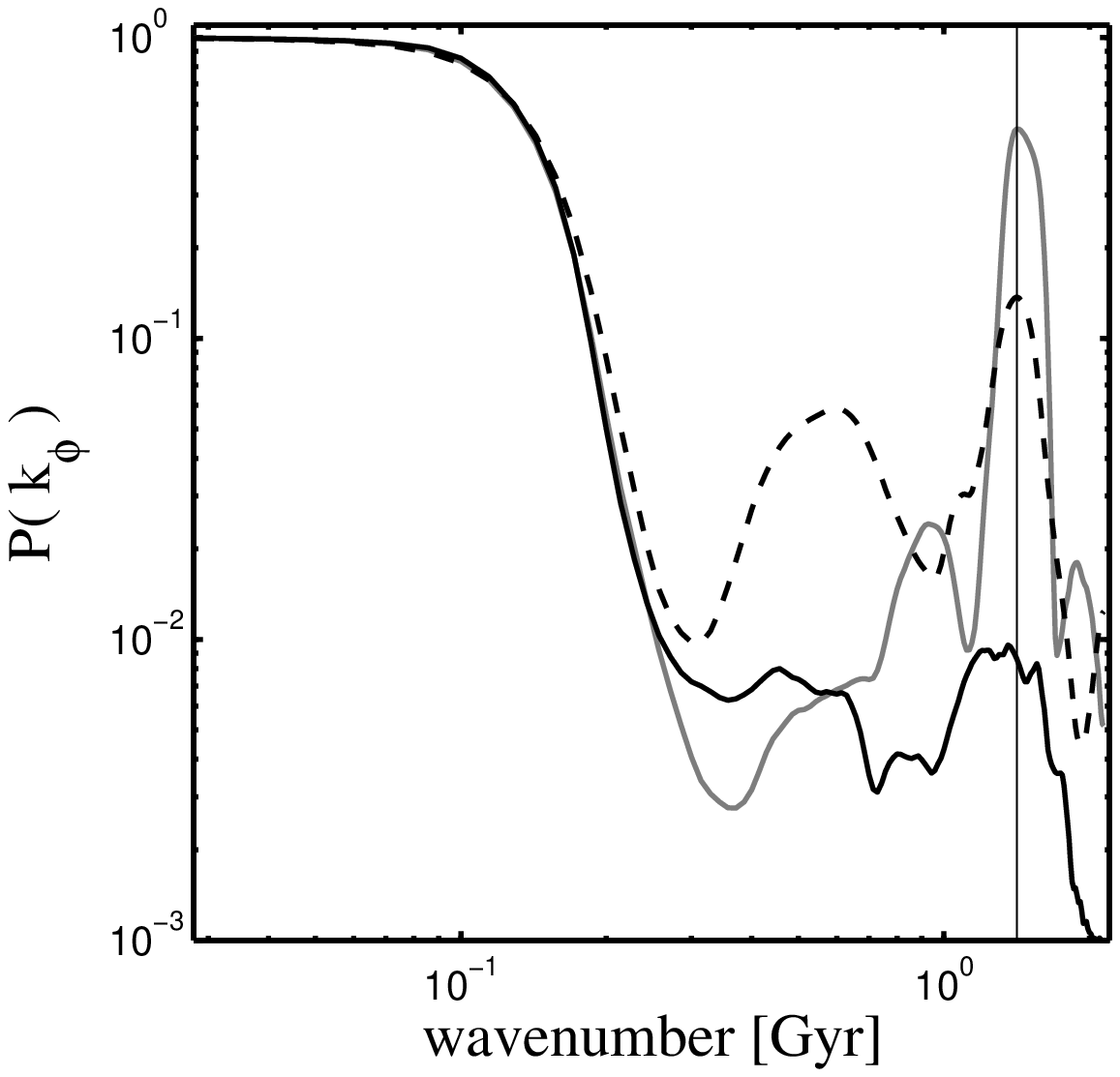}

\caption{Top  panels: Distribution of  stellar particles  in frequency
  space for  the three satellites  shown in Fig.~8,  after convolution
  with  the  {\it Gaia}  observational  errors.   Grey  dots show  the
  distribution  of  all  the   stellar  particles  inside  the  `Solar
  Neighbourhood'  sphere, whereas  the red  dots are  the  subset with
  $\sigma_{\varpi}  / \varpi  \le  0.02$ (see  also zoom-in).   Bottom
  panels: The black solid lines  show the 1-D normalized power spectra
  obtained from  the distribution of  points shown in the  top panels,
  while  the  black  dashed  lines  corresponds  to  the  subset  with
  $\sigma_{\varpi}/\varpi \leq  0.02$.  The grey solid  lines show the
  normalized power spectra obtained from the distribution of particles
  before     convolution     with     the     observational     errors
  (cf.~Figure~\ref{fig:sats_frq}).   Their largest  amplitude  peak is
  indicated whith a vertical line.}
\label{fig:sats_frq_err}
\end{figure*}

A  derivation  of  the  time  since accretion  is  significantly  more
difficult  when  the {\it  Gaia}  measurement  errors  are taken  into
account.   We  exemplify this  by  focusing  on  the three  satellites
described in  Section~\ref{sec:time_acc}.  The `observed' distribution
of stellar particles  in frequency space are shown in  grey in the top
panels  of  Figure~\ref{fig:sats_frq_err}.   A  direct  comparison  to
Figure~\ref{fig:sats_frq} clearly  highlights what the  effects of the
{\it Gaia}  errors are.  As in  Section~\ref{sec:time_acc}, we perform
our Fourier  analysis and compute the normalized  power spectra. These
are  shown  with   a  black  solid  line  in   the  bottom  panels  of
Figure~\ref{fig:sats_frq_err}.   Only for  the spectrum  shown  in the
middle  panel, we  can  identify a  single  dominant peak.   Therefore
observational  errors seem  to be  large  enough to  erase the  signal
associated to  the time since accretion  in the power  spectrum for at
least some of our satellite galaxies.

We wish  to obtain  an order  of magnitude estimate  of how  large the
errors  in velocities  have to  be to  blend two  adjacent  streams in
frequency space.   Let us  consider a satellite  moving on  a circular
orbit  accreted  $t_{\rm acc}=  8$~Gyr  ago.   The  separation of  two
adjacent streams in frequency space  is $\Delta \Omega = 2 \pi/ t_{\rm
acc} \approx 0.78$ (comparable to that found for the satellites in our
simulations).  Since  $V_{\phi} =  R \Omega$, we  may deduce  that the
maximum  error  in  the  tangential velocity  should  be  $\sigma_{\rm
v_{\phi}} = R \Delta {\Omega}$. Therefore, for the streams found at $R
= 8$~kpc from  the Galactic centre, $\sigma_{\rm v_{\phi}}  \leq 6$ km
s$^{-1}$. This  implies that,  to be able  to estimate the  time since
accretion from the power  spectra, the relative parallax errors should
be $\sigma_{\varpi} / \varpi \leq 0.02$\footnote{In this derivation we
have assumed  that relative  errors in proper  motion are of  the same
order of magnitude than those  in the parallax.  However, for the {\it
Gaia} mission  the former are  expected to be  in general an  order of
magnitude smaller.}.

The  set  of stellar  particles  from  each  of our  three  satellites
satisfying this  condition are shown as  red dots in the  top panel of
Figure~\ref{fig:sats_frq_err}.   Out  of the  445  (left panel),  1264
(middle  panel) and  720  (right panel)  stellar particles  originally
found  inside our  solar neighbourhood  sphere, only  90, 239  and 113
respectively, have remained after the error cut.  The normalized power
spectra obtained  for these distributions are shown  with dashed black
lines  in bottom  panels of  Figure~\ref{fig:sats_frq_err}.   Now, the
largest  amplitude  peak  is  (once  again  and  for  all  satellites)
associated to the time since accretion.  Note that, since each peak is
located  at   the  same  wavenumber   as  in  the  analysis   with  no
observational errors (grey curves in this Figure), the estimated times
since  accretion for  each  satellite are  exactly  those obtained  in
Section~\ref{sec:time_acc}.

Nevertheless, it is  clear that the signal found  in the power spectra
could  be  determined  better  if   a  larger  number  of  stars  with
$\sigma_{\varpi}/\varpi \leq 0.02$ could  be available. Such stars are
expected to be present in the {\it Gaia} catalogue, but because of the
limited numerical resolution of our  experiments, they are not part of
our  Mock {\it  Gaia} catalogue.   Such  stars would  be fainter  than
$M_{\rm V}  =4.5$, but  closer than 2~kpc  from the Sun.  For example,
according to the  latest {\it Gaia} performance numbers,  a dwarf star
located  at  1~kpc from  the  Sun, with  an  apparent  magnitude of  V
$\approx$ 16 should have a parallax measurement error $\sigma_{\varpi}
/ \varpi  \leq 0.02$. This apparent magnitude  and distance translates
into an  absolute magnitude of $M_{\rm  V} \approx 6$, i.e.  it is 1.5
magnitudes  fainter  than  limiting  magnitude  we have  used  in  our
simulations.  If  we use  the number of  particles found  at distances
closer  than 1~kpc  from the  Sun,  and if  we take  into account  the
luminosity functions of the 3  satellites in Fig.~10, we can obtain an
estimate of how many stars would  be observable by {\it Gaia} with the
desired parallax errors. In this  way we find, respectively, 186, 465,
and 276  extra stars with  $4.5 \leq M_{V}  \leq 12$, which  will thus
allows to estimate the time since accretion for each satellite.

\subsection{Identification of Satellites}
Although promising, this is unlikely to  be the way we will proceed in
the future  with real {\it Gaia}  observations. In order  to obtain an
estimate of the time since  accretion of any given satellite, we first
have to efficiently identify it.   This can be performed by applying a
suitable clustering  technique.  In  this work we  have used  the Mean
Shift algorithm \citep{fukuhost,comameer,derpa}.  The main idea behind
mean shift is  to treat the points in any  $N$-dimensional space as an
empirical  probability density  function  where dense  regions in  the
space correspond  to the local maxima of  the underlying distribution.
For  each data  point in  the space,  one performs  a  gradient ascent
procedure on the local estimated density until convergence is reached.
Furthermore, the data points associated with the same stationary point
are considered members of the same cluster.

We  have applied  the Mean  Shift algorithm  to the  set  of particles
inside a  sphere of 4~kpc radius,  located at 8~kpc  from the galactic
centre, projected into  the space of $E-L-L_{z}$. We  chose this space
because the satellite's internal  substructure (due to the presence of
individual  streams) is  less  well defined  and,  therefore, is  more
suitable for  a clustering search  of global structures.   Enclosed in
this  sphere  we  find a  total  of  $\sim  8 \times  10^{4}$  stellar
particles, coming from 26  different satellites contributing each with
at least 20 particles.  In addition, the disc and the bulge contribute
with $\sim  20\ 000$ and  $\sim 2\ 000$ particles,  respectively.  For
this  analysis  we especially  chose  to  sample  a larger  volume  of
physical space  so that  each satellite is  represented with  a larger
number   of  stellar   particles.    In  this   way,   we  can   avoid
overfragmentation, which occurs when a  clump is populated with a very
small number of particles.

\begin{figure}
\centering
\includegraphics[width=83mm,clip]{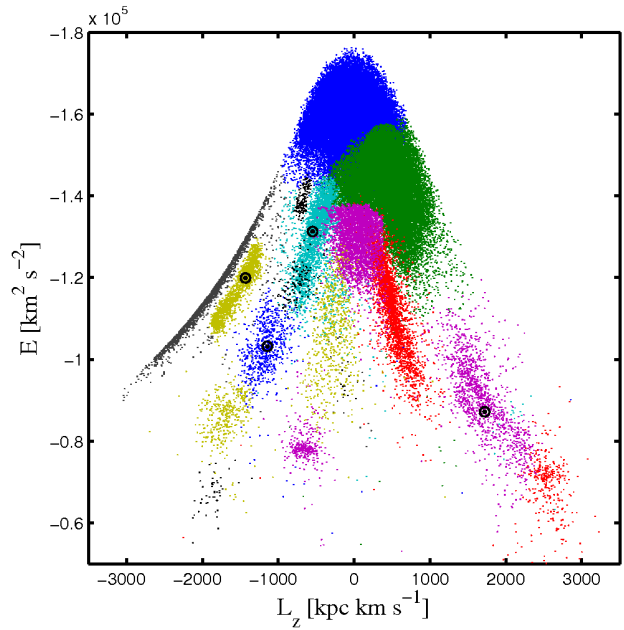}
\caption{Distribution of  stellar particles  inside a sphere  of 4~kpc
  radius at 8~kpc from the galactic centre in $E$ vs. $L_{z}$ space as
  would  be observed  by {\it  Gaia}. The  different colours  show the
  groups identified  by the Mean  Shift algorithm. Black  open circles
  denote  those for which  the time  since accretion  was successfully
  derived.}
\label{fig:mstest}
\end{figure}

Figure~\ref{fig:mstest}  shows  the  different clusters  of  particles
identified with  this method.   We have found  17 groups  that contain
more than  20 particles. Out these  17 groups, only 15  have more than
$50\%$ of  its particles  associated to a  single progenitor.   One of
these  groups corresponds  to  the  disc while  two  other are  double
detections  of two different  satellites that  were fragmented  by the
algorithm.   Therefore,  only  12  of  these groups  can  be  uniquely
associated to a  single satellite. This corresponds to  $\sim 50\%$ of
all the satellites contributing with  stars to this volume.  This is a
similar recovery rate to that  obtained by \citet{hz00}, but now under
a more realistic cosmological model  and with the latest model for the
{\it Gaia} measurement errors.

When  attempting  to  compute   the  time  since  accretion,  we  were
successful  only  in  four  cases.   These  groups  are  indicated  in
Figure~\ref{fig:mstest}  with   black  open  circles.    Two  of  them
correspond  to the  satellites  labeled  number 1  and  2 in  previous
analysis.  In  general, we find  that the remaining  identified groups
lack  a  significant number  of  `stars'  with  the required  relative
parallax error, $\sigma_{\varpi} /  \varpi \leq 0.02$ (i.e., typically
$\leq  50$).  However,  as explained  before, in  this  simulation our
limited numerical resolution led us to consider only stars with $M_{V}
\leq 4.5$.   After estimating the  number of stellar  particles within
1~kpc  from the  Sun  that may  be  observed by  {\it  Gaia} with  the
required    relative     parallax    errors    (as     explained    in
Section~\ref{sec:est_time_acc_err}),  we   find  that  at   least  two
additional satellites,  among the 12 previously  isolated, should have
at least 200  stellar particles available to compute  a reliable power
spectrum.

\section{Summary and Conclusions}

We  have studied  the characteristics  of merger  debris in  the Solar
neighbourhood as  may be observed by  ESA's {\it Gaia}  mission in the
near  future. We  have  run a  suite  of $N$-body  simulations of  the
formation  of the  Milky Way  stellar  halo set  up to  match, at  the
present  time, its known  properties such  as the  velocity ellipsoid,
density  profile and  total  luminosity.  The  simulations follow  the
accretion of a  set of 42 satellite galaxies  onto a semi-cosmological
time dependent  Galactic potential.  These satellites  are evolved for
10~Gyr,  and  we  use  the  final  positions  and  velocities  of  the
constituent particles to generate a Mock {\it Gaia} catalogue.

Using synthetic  CMDs, we have resampled the  satellite's particles to
represent  stars down to  $M_V \approx  4.5$. This  absolute magnitude
corresponds to an apparent magnitude  $V \approx 16$ at 2.5~kpc, which
is  comparable  to the  {\it  Gaia}  magnitude  limit for  which  full
phase-space information  will be  available.  Our Mock  catalogue also
includes a  Galactic background population  of stars represented  by a
Monte   Carlo  model   of  the   Galactic  disc   and  bulge,   as  in
\citet{brown05}.  At 8~kpc from  the Galactic  centre, stars  from the
disc  largely outnumber  those  of  the stellar  halo,  however it  is
possible  to  reduce  their  impact   by  applying  a  simple  cut  on
metallicity.  We  have estimated,  using the latest  determinations of
the MDF of the Galactic  disc(s) \citep{gc,ive08}, that only $10\ 000$
stars out  of the  estimated $4.1 \times  10^{7}$ disc  stars brighter
than $V=17$ in the Solar neighbourhood should have [${\rm Fe}/{\rm H}]
\leq -1.1$.  A smaller number  of bulge stars ($\sim 800$) with [${\rm
Fe}/{\rm H}]  \leq -1.1$ is  expected to contaminate our  stellar halo
sample,  down to  the faintest  $M_V$. This  fraction  represents only
$2.6$ per  cent of the whole  Mock {\it Gaia}  stellar halo catalogue,
and therefore does not constitute an important obstacle to our ability
to characterize this component.

Finally, we have convolved the positions and velocities of all `observable
stars' in our Solar Neighbourhood sphere with the latest models of the
{\it Gaia} observational errors, according to performances given by
ESA (http://www.rssd.esa.int/gaia).

The  analysis presented  here confirms  previous results,  namely that
satellites can  be identified as coherent clumps  in phase-space, e.g.
in  the $E$-$L_z$  projection \citep[see][]{hz00,  knebe05,  font}. We
find    that    a   clustering    algorithm    such   as    Mean-Shift
\citep[][]{fukuhost,comameer,derpa} is able to recover roughly 50\% of
all   satellites   contributing  stellar   particles   to  the   Solar
neighbourhood sphere.

We have also demonstrated that  even after accounting for the Galactic
background contamination  and the expected measurement  errors for the
{\it Gaia} mission, the space of orbital frequencies is also very rich
in substructure. In this space streams from a given satellite define a
regular  pattern  whose  characteristic  scale is  determined  by  the
satellite's time  since accretion.  We find  that reasonable estimates
of the time  since accretion may be provided when  the number of stars
with accurate parallaxes ($\sigma_{\varpi} / \varpi \leq 0.02$) from a
given  satellite is at  least $\sim  100$.  This  was possible  in our
simulations for 4 cases.

Together  with  {\it  Gaia},  ground  based  follow-up  campaigns  are
currently  being planned  by European  astronomical  community.  Their
purpose  is   to  complement  a  future  {\it   Gaia}  catalogue  with
information that either will not be obtained or for which the accuracy
will  be  very low.   As  an  example,  high resolution  multi  object
spectroscopy  in combination  with ground  based photometry  could not
only push the limiting magnitude  of the phase-space catalogue down to
$V  \approx 20$, but  could also  provide detailed  chemical abundance
patterns of individual stars.  The former will require the development
of accurate and precise  photometric distances indicators which can be
calibrated  using the large  number of  stars near  the Sun  with very
precise trigonometric parallaxes  from {\it Gaia}.  The identification
of satellites in, e.g., $E-L_{z}$ space could be considerably improved
by having such an  extended full phase-space catalogue.  Firstly, very
faint satellites could now be  observed and, secondly, the much larger
sample of stars could allow us to apply a clustering algorithm in very
small volumes  around the  Sun with enough  resolution to  avoid large
overfragmentation.  The  later is  important because by  extending the
volume probed  we expect to  reduce the overlap between  satellites in
$E-L_{z}$ space.  In addition, chemical tagging will be very important
to characterize the satellite's  star formation and chemical histories
\citep{FBH}.   Note, however,  that in  our simulations  we  find that
stellar particles in a given  stream do not originate from a localized
region  in  physical  space   (such  as  a  single  molecular  cloud).
Therefore even individual  streams are likely to reflect  the full MDF
present in the object at the time it was accreted.

Although  our satellites  were evolved  in a  cosmologically motivated
time  dependent host  potential, our  simulations do  not  contain the
fully   hierarchical   and  often   chaotic   build-up  of   structure
characteristic of  the $\Lambda$ cold dark matter  model.  The violent
variation  of the  host potential  during merger  events,  the chaotic
behavior induced by a triaxial dark matter halo \citep[e.g.][]{vol08},
and the orbital evolution  due to baryonic condensation \citep{vallu},
are potentially  important effects which we have  neglected and should
be taken into account in future work.  To assess the impact of some of
these effects on the distribution of debris in the Solar Neighbourhood
we  are  currently  analysing  a fully  cosmological  high  resolution
simulation of the formation of  the Galactic stellar halo based on the
Aquarius project \citep{cooper}.

\section*{Acknowledgments}
We are  very grateful  to Daniel Carpintero  for the software  for the
spectral  analysis.  AH  acknowledges the  financial support  from the
European Research Council under ERC-StG grant GALACTICA-240271 and the
Netherlands Organization for Scientific  Research (NWO) through a VIDI
grant.  AH and  Y-SL acknowledge  the financial  support from  the NWO
STARE program  643.200.501.  This  work has made  use of  the IAC-STAR
Synthetic CMD  computation code.  IAC-STAR is  supported an maintained
by the computer division of the IAC.

\label{lastpage}

\begin{thebibliography}{aa}

\bibitem[Aparicio \& Gallart(2004)]{apagalla}Aparicio A., Gallart C., 2004, AJ, 128, 1465
\bibitem[Arifyanto \& Fuchs(2006)]{ari_fu} Arifyanto M.~I., Fuchs B., 2006, A\&A, 449, 533

\bibitem[Belokurov et al.(2007)]{belu07} Belokurov V. et al., 2007, ApJ, 658, 337
\bibitem[Belokurov et al.(2006)]{belu06} Belokurov V. et al., 2006, ApJ, 642, L137
\bibitem[Brown, Vel\'asquez \& Aguilar(2005)]{brown05} Brown A.~G.~A., Vel\'asquez H.~M., Aguilar L.~A., 2005, MNRAS, 359, 287
\bibitem[Bryan \& Norman(1998)]{bn} Bryan G., Norman M., 1998, ApJ, 495, 80
\bibitem[Bullock \& Johnston(2005)]{bj05} Bullock J.~S., Johnston K.~V., 2005, ApJ, 635, 931

\bibitem[Carpintero \& Aguilar(1998)]{daniel} Carpintero D.~D., Aguilar L.~A., 1998, MNRAS, 298, 1
\bibitem[Comaniciu \& Meer(2002)]{comameer} Comaniciu D., Meer P., 2002, IEEE Transactions on Pattern Analysis and Machine Intelligence, 25(5), 564–577
\bibitem[Cooper et al.(2010)]{cooper} Cooper A.~P. et al., 2010, MNRAS in press (arXiv:0910.3211)

\bibitem[De Lucia \& Helmi(2008)]{lh} De Lucia G., Helmi A., 2008, MNRAS, 391, 14
\bibitem[Derpanis(2005)]{derpa} Derpanis K.~G., 2005, \href{http://www.cse.yorku.ca/~kosta/CompVis_Notes/mean_shift.pdf}{weblink}


\bibitem[Font et al.(2006)]{font} Font A.~S., Johnston K.~V., Bullock J.~S, Robertson B.~E., 2006, ApJ, 646, 886
\bibitem[Fuchs \& Jahrei{\ss}(1998)]{fuchs} Fuchs B., Jahrei{\ss} H., 1998, A\&A, 329, 81
\bibitem[Fukunaga \& Hostetler(1975)]{fukuhost} Fukunaga K., Hostetler L., 1975, IEEE Transactions on Information Theory, 21(1), 32–40
\bibitem[Freeman \& Bland-Hawthorn(2002)]{FBH} Freeman K., Bland-Hawthorn J., 2002, ARA\&A, 40, 487

\bibitem[G\'{o}mez \& Helmi(2010)]{gh10} G\'{o}mez F.~A., Helmi A., 2010, MNRAS, 401, 2285
\bibitem[Grillmair(2006)]{grill} Grillmair C.~J., 2006, ApJ, 645, L37
\bibitem[Guzm\'an, Lucey \& Bower(1993)]{guzman}Guzm\'an R., Lucey J.~R., Bower R.G., 1993, MNRAS, 265, 731

\bibitem[Helmi(2008)]{hreview} Helmi A., 2008, A\&AR, 15, 145
\bibitem[Helmi \& de Zeeuw(2000)]{hz00} Helmi A., de Zeeuw P.~T., 2000, MNRAS, 319, 657
\bibitem[Helmi \& White(1999)]{hw} Helmi A., White S.~D.~M., 1999, MNRAS, 307, 495
\bibitem[Helmi et al.(1999)]{hwzz99} Helmi A., White S.~D.~M., de Zeeuw P.~T., Zhao H., 1999, Nature, 402, 53
\bibitem[Helmi, White \& Springel(2003)]{hws03} Helmi A., White S.~D.~M., Springel V., 2003, MNRAS, 339, 834
\bibitem[Helmi et al.(2006)]{h06} Helmi A., Navarro J.~F., Nordstr\"om B., Holmberg J., Abadi M.~G.,  Steinmetz M., 2006, MNRAS, 365, 1309
\bibitem[Hernquist(1990)]{hernq} Hernquist L., 1990, ApJ, 356, 359
\bibitem[H{\o}g et al.(2000)]{thyco} H{\o}g E., Fabricius C., Makarov V.~V. et al., 2000, A\&A, 355, L27

\bibitem[Ibata et al.(2003)]{ibata03} Ibata R., Irwin M.~J., Lewis G.~F., Ferguson A.~M.~N., Tanvir N., 2003, MNRAS, 340, L21
\bibitem[Ibata et al.(2001a)]{ibata01a}Ibata R.~A., Irwin M.~J., Lewis G.~F., Ferguson A.~M.~N., Tanvir N., 2001a, Nature, 412, 49
\bibitem[Ibata et al.(2001b)]{ibata01b} Ibata R.~A., Irwin M.~J., Lewis G.~F., Stolte A., 2001b, ApJ, 547, L133
\bibitem[Ibata, Gilmore \& Irwin(1994)]{ibata94} Ibata R.~A., Gilmore G., Irwin M.~J., 1994, Nature, 370, 194
\bibitem[Ivezi\'{c} et al.(2008)]{ive08} Ivezi\'{c} \u{Z}. et al., 2008, ApJ, 684, 287

\bibitem[Johnston et al.(2008)]{johns08} Johnston K.~V., Bullock J.~S., Sharma S., Font A., Robertson B.~E., Leitner S.~N., 2008, ApJ, 689, 936

\bibitem[King(1966)]{king} King I.~R., 1966, AJ, 71, 64
\bibitem[Klement et al.(2009)]{klement09} Klement R. et al., 2009, ApJ, 698, 865
\bibitem[Knebe et al.(2005)]{knebe05} Knebe A., Gill S., Kawata D., Gibson B.~K., 2005, MNRAS, 357, 35
\bibitem[Koposov et al.(2008)]{kopo} Koposov S. et al., 2008, ApJ, 686, 279
\bibitem[Kroupa et al.(1993)]{kroupa} Kroupa P., Tout C.~A., Gilmore G., 1993, MNRAS, 262, 545

\bibitem[Majewski et al.(2003)]{majewski} Majewski S.~R., Skrutskie M.~F., Weinberg M.~D., Ostheimer J.~C., 2003, ApJ, 599, 1082 
\bibitem[Marigo et al.(2008)]{marigo} Marigo P., Girardi L., Bressan A., Groenewegen M.~A.~T., Silva, L., Granato G.~L., 2008, A\&A, 482, 883
\bibitem[Mart\'inez-Delgado et al.(2009)]{Mart09} Mart\'inez-Delgado D., Pohlen M., Gabany R.J., Majewski S.R., Pe\~narrubia, J., Palma C., 2009, ApJ, 692, 955
\bibitem[Mart\'inez-Delgado et al.(2008)]{Mart08} Mart\'inez-Delgado D., et al., 2008, ApJ, 689, 184
\bibitem[McConnachie et al.(2009)]{McCo09} McConnachie A.~W. et al., 2009, Nature, 461, 6673
\bibitem[McMillan \& Binney(2008)]{mcm} McMillan P.~J., Binney J., 2008, MNRAS, 390, 429 
\bibitem[Miyamoto \& Nagai(1975)]{mi-na} Miyamoto M., Nagai R., 1975, PASJ, 27, 533
\bibitem[Mihalas \& Binney(1981)]{mihabin}Mihalas D., Binney J., 1981, Galactic Astronomy: Structure and Kinematics, 2nd ed., New York NY, W.H. Freeman and Company

\bibitem[Navarro, Frenk \& White(1996)]{nfw} Navarro J.~F., Frenk C.~S., White S.~D.~M., 1996, ApJ, 462, 563
\bibitem[Newberg et al.(2002)]{new02} Newberg H.~J. et al., 2002, ApJ, 569, 245
\bibitem[Nordstr\"{o}m et al.(2004)]{gc} Nordstr\"{o}m B. et al., 2004, A\&A, 418, 989

\bibitem[Perryman et al.(2001)]{gaia} Perryman M.~A.~C. et al., 2001, A\&A, 369, 339
\bibitem[Perryman et al.(1997)]{hip} Perryman M.~A.~C. et al., 1997, A\&A, 323, 49
\bibitem[Pietrinferni et al.(2004)]{pietri} Pietrinferni A., Cassisi S., Salaris M., Castelli F., 2004, ApJ, 612, 168
\bibitem[Plummer(1911)]{plum} Plummer H.~C., 1911, MNRAS, 71, 460

\bibitem[Schuster \& Allen(1997)]{schall} Schuster W.~J., Allen C., 1997, A\&A, 319, 796
\bibitem[Sharma \& Johnston(2009)]{sj09} Sharma S., Johnston K.~V., 2009, ApJ, 703, 1061
\bibitem[Smith et al.(2009)]{smith09} Smith M.~C., Evans N.~W, Belokurov V., Hewett P.~C., Bramich D.~M., Gilmore G., Irwin M.~J., Vidrih S., Zucker D.~B., 2009, MNRAS, 399, 1223
\bibitem[Smith et al.(2007)]{smith07} Smith M.~C. et al., 2007, MNRAS, 379, 755
\bibitem[Sofue, Honma \& Omodaka(2009)]{sofue} Sofue Y., Honma M., Omodaka T., 2009, PASJ, 61, 227
\bibitem[Springel(2005)]{springel2005} Springel V., 2005, MNRAS, 364, 1105
\bibitem[Starkenburg et al.(2009)]{else09} Starkenburg E. et al., 2009, ApJ, 698, 567

\bibitem[Tumlinson(2010)]{tumli} Tumlinson J., 2010, ApJ, 708, 1398

\bibitem[Yanny et al.(2009)]{yanny09} Yanny B. et al., 2009, ApJ, 137, 4377
\bibitem[Yanny et al.(2003)]{yanny03} Yanny B. et al., 2003, ApJ, 588, 824

\bibitem[Valluri et al.(2010)]{vallu} Valluri M., Debattista V.~P., Quinn T., Moore B., 2010, MNRAS, 403, 525
\bibitem[Vogelsberger et al.(2008)]{vol08} Vogelsberger M., White S.~D.~M., Helmi A., Springel V., 2008, MNRAS, 385, 236

\bibitem[Wechsler et al.(2002)]{wech}Wechsler R.~H., Bullock J.~S., Primack J.~R., Kravtsov A.~V., Dekel A., 2002, ApJ, 568, 52
\bibitem[White \& Rees(1978)]{wr} White S.~D.~M., Rees M.~J., 1978, MNRAS, 183, 341
\bibitem[White \& Springel(2000)]{ws2000} White S.~D.~M., Springel V., 2000, in The First Stars, Weiss A., Abel T.~G., Hill V., eds., p. 327

\bibitem[Zhao et al.(2003)]{zhao} Zhao D.~H., Mo H.~J., Jing Y.~P., B\"orner G., 2003, MNRAS, 339, 12
\bibitem[Zoccali(2009)]{zoca} Zoccali M., 2009, arXiv0910.5133Z
\bibitem[Zwitter et al.(2008)]{zwitter08} Zwitter T. et al., 2008, AJ, 136, 421

\end{thebibliography}
\end{document}